\documentclass[12pt]{article}

\usepackage[utf8]{inputenc}
\usepackage[T1]{fontenc}
\usepackage[a4paper,margin=1in]{geometry}
\usepackage{amsmath,amssymb,amsthm}
\usepackage{graphicx}
\usepackage{booktabs}
\usepackage{threeparttable}
\usepackage{rotating}
\usepackage{multirow}
\usepackage{makecell}
\usepackage{array}
\usepackage{float}
\usepackage{comment}
\usepackage[section]{placeins}
\usepackage{textcomp}
\usepackage{xcolor}
\usepackage[
  colorlinks=true,
  linkcolor=blue,
  citecolor=blue,
  urlcolor=blue
]{hyperref}
\usepackage[authoryear]{natbib}

\newcommand{\bodyspacing}{\linespread{1.5}\selectfont}
\AtBeginEnvironment{table}{\linespread{1}\selectfont}
\AtBeginEnvironment{sidewaystable}{\linespread{1}\selectfont}
\AtBeginEnvironment{figure}{\linespread{1}\selectfont}

\theoremstyle{plain}

\newtheorem{assumption}{Assumption}

\theoremstyle{definition}
\newtheorem{definition}{Definition}

\theoremstyle{remark}

\newcommand{\E}{\mathbb{E}}

\DeclareMathOperator*{\argmax}{arg\,max}

\definecolor{scolor}{RGB}{0,0,0}
\newcommand{\anchoredtablecaption}[2]{%
  \refstepcounter{table}%
  \par\centering{\small\textbf{Table \thetable}\par#1\par}\label{#2}\vspace{6pt}%
}
\newcommand{\anchoredfigurecaption}[2]{%
  \refstepcounter{figure}%
  \par\centering{\small\textbf{Figure \thefigure:}~#1\par}\label{#2}\vspace{6pt}%
}

\title{
\textbf{Daycare Matching with Siblings: Social Implementation and Welfare Evaluation}
\thanks{
We thank Koriyama City for providing the data and for the social implementation of the priority reform.
We are grateful for helpful feedback from Yu Awaya, Kenzo Imamura, Yuichiro Kamada, Michihiro Kandori, Fuhito Kojima, Tatsushi Oka, Josué Ortega, Fumio Otake, Yuta Toyama, as well as participants at SWET 2025, RIETI, ACM EC 2026, and EARIE 2026.
This work was supported by JST ERATO Grant Number JPMJER2301, Japan.
The views expressed in this paper are those of the authors and do not necessarily represent Koriyama City, CyberAgent, or the University of Tokyo Market Design Center.
}
}

\author{
Kan Kuno\thanks{The University of Tokyo. Email: \texttt{kankuno@e.u-tokyo.ac.jp}}
\and
Daisuke Moriwaki\thanks{CyberAgent, Inc. Email: \texttt{moriwaki\_daisuke@cyberagent.co.jp}}
\and
Yoshihiro Takenami\thanks{CyberAgent, Inc. Email: \texttt{takenami\_yoshihiro@cyberagent.co.jp}}
}

\date{September 2026}

\begin{document}

\begin{titlepage}
\maketitle
\begin{abstract}
\noindent
In centralized matching markets, agents may value joint assignment, as with siblings or couples. Standard preference estimation ignores such complementarities, complicating welfare analysis of priority rules for paired assignment. We develop an empirical framework incorporating these preferences and apply it to Japanese daycare assignment. Families face both additional commuting distance and a fixed disutility from split assignment. We estimate the latter at 4.61 commuting-kilometer equivalents. Our fixed-report counterfactual estimates that the reform increased mean welfare by 0.032 kilometer-equivalent units. Ignoring sibling complementarity understates welfare gains for households applying simultaneously for multiple children by about 27\%.

\vspace{0.5em}
\noindent\textbf{Keywords:} market design, demand complementarities, daycare assignment, sibling priority, structural estimation

\noindent\textbf{JEL Classification:} D47, I21, D12
\end{abstract}
\setcounter{page}{0}
\thispagestyle{empty}
\end{titlepage}

\pagebreak
\newpage

\bodyspacing

\section{Introduction}

In centralized assignment mechanisms, families with multiple children may value having their children assigned to the same facility. In school choice, placing children in different schools can impose substantial logistical costs, making joint assignment a first-order concern and motivating sibling priority in many mechanisms. At the same time, such preferential treatment raises equity concerns for families with only one child. In the United Kingdom, for example, the national school admissions watchdog expressed concerns in 2013 that granting top priority to siblings could disadvantage first-born children.\footnote{Richard Garner, ``Schools urged to scrap `unfair' admissions rule for siblings,'' \emph{The Independent}, November 29, 2013, \url{https://www.independent.co.uk/news/education/education-news/schools-urged-to-scrap-unfair-admissions-rule-for-siblings-8973650.html}.} This illustrates a broader pattern in which assignment mechanisms that prioritize joint assignments, such as sibling priority, must balance the efficiency gains from accommodating such preferences against the resulting equity concerns.

To understand this efficiency--equity tradeoff, an empirical framework is needed that explicitly accounts for siblings' joint preferences. Such preferences create a complementarity: the value of one child's assignment depends on where the other child is assigned, because placing them at different facilities imposes a logistical cost. However, standard methods for preference estimation in school choice treat students' choices separately and do not model household preferences over joint assignments. We propose an estimation method that accommodates sibling complementarities and apply it to the Japanese daycare context. Sibling priority is also a salient issue in this setting, where daycare assignment is centralized and operates in a manner similar to school choice. For example, a parent commented to a municipality: "My children attend different facilities, and picking up and dropping off are far more difficult than expected. Each facility has different supplies and communication methods."\footnote{Kawasaki City, Child and Future Bureau, ``Results of the Public Comment Procedure on Expanding Support for Households with Multiple Children in the Use of Childcare Facilities,'' August 30, 2023, comment 3, \url{https://www.city.kawasaki.jp/980/cmsfiles/contents/0000151/151291/20230830-3(5).pdf} (authors' translation).} Our estimation framework allows us to quantify these burdens.

In our model, parents with multiple children have preferences over tuples of assignments. As in \citet{gentzkow_valuing_2007}, utility includes a complementarity across individual choices: the family receives the sum of each child's assignment utility but incurs a fixed penalty and additional trip-chain commuting distance if the children attend different facilities.
We follow \citet{fack_beyond_2019} in converting the stability property of the observed match into a discrete choice problem. To account for sibling preferences, we use the extended stability notion of \citet{sun_stable_2024}, which allows for preferences over sibling assignments.
We also use reported-preference data as moment inequalities to help distinguish household-level facility tastes from the cost of split assignment.

We use data from Koriyama City in Fukushima Prefecture, located in northeastern mainland Japan. Before the reform, Koriyama gave no sibling priority to children applying together. Many parents strongly preferred joint non-assignment to split assignment, so the lack of sibling priority for simultaneous applicants affected their access to acceptable joint placements. Japanese daycare assignment relies on a single-index priority score, and the policy question facing the municipality was how much weight to assign to sibling applicants, as excessive weight could give rise to the equity concerns discussed above.
Koriyama implemented a priority reform to make joint placement easier for siblings, with the point values selected based on our simulated outcomes under alternative priority combinations.

We use the municipality's post-reform application data to estimate applicant preferences. In addition to the expected findings that applicants dislike longer commuting distances and that parents with higher opportunity costs of parental home care, such as full-time working mothers, derive higher utility from daycare services, we find substantial disutility associated with split assignment. Beyond the additional commuting distance it entails, split assignment generates a non-distance disutility equivalent to 4.61 kilometers of commuting for the reference household, about twice the mean positive commuting distance in the sample. This additional cost likely reflects reduced family coordination, differences in daily schedules, and uncoordinated communication across facilities. A canonical model that ignores sibling complementarity not only omits the fixed cost of split assignment but also estimates a distance penalty of 0.817, about 17\% below the 0.981 estimate in the baseline model. It treats sibling applicants who prefer joint assignment at more distant daycares over split assignment at closer daycares as if they simply have weaker distance sensitivity.

We simulate counterfactual assignments under alternative weights on sibling priority for the fixed 2025 applicant pool and evaluate welfare conditional on submitted preferences. The simulated sibling-priority change raises mean welfare by 0.032 kilometer-equivalent units, with gains concentrated among households applying simultaneously for multiple children. These households gain 1.214 km on average, mainly from the sum of children's individual assignment utilities (1.130 km), with lower split-assignment costs contributing another 0.039 km. Households applying for a child without siblings lose 0.089 km. The canonical model estimates a gain of 0.883 km for households applying simultaneously for multiple children, about 27\% less than the baseline model: it cannot capture the reduction in split-assignment costs and also estimates a smaller gain in children's individual assignment utilities. We also find that a welfare-maximizing point grants substantially more priority to both sibling groups than the actual priority reform. Compared with the simulated post-reform mix, joint assignment among households applying simultaneously for multiple children rises from 76.6\% to 85.2\%, but the assignment rate for households applying for a child without siblings falls from 77.1\% to 75.6\%, illustrating the efficiency--equity tradeoff.

\section{Related Literature}

This paper contributes to the literature on preference estimation in centralized matching models with non-transferable utility. Existing work primarily estimates individual student preferences over schools, including \citet{agarwal_demand_2018}, \citet{calsamiglia_priorities_2018}, and \citet{fack_beyond_2019}, as well as related work for daycare markets such as \citet{kuno_dynamic_2023}. 
While \citet{agarwal_demand_2018} and \citet{calsamiglia_structural_2020} estimate the value to an applying student of being assigned to the school attended by an older sibling, they take the older sibling's assignment as given.
We instead explicitly model household preferences over the pair of sibling assignments and estimate the value of placing them together.
Methodologically, our approach is closest to \citet{fack_beyond_2019}, who exploit the stability property of the matching outcome and translate it into a discrete choice framework. 
We combine ideas from demand estimation for bundled goods, as in \citet{gentzkow_valuing_2007}, with school choice matching mechanisms.

This paper relates to the literature on matching with siblings and couples. Early work by \citet{roth1984evolution} shows that stable matchings may fail to exist when couples have joint preferences, and \citet{ronn1990np} shows that verifying existence is computationally hard. Despite these theoretical difficulties, \citet{roth1999redesign} document that stable outcomes frequently arise in practice. Building on this motivation, \citet{klaus2005stable} establish existence under weakly responsive preferences, and \citet{kojima2013matching} and \citet{ashlagi2014stability} establish existence results for large markets. In school choice settings, \citet{dur_family_2022} study mechanisms that accommodate family considerations, and \citet{correa_school_2019} document the large-scale implementation of a nationwide school assignment system in Chile that favors joint sibling assignment. Focusing on Japanese daycare markets, \citet{sun_stable_2024} propose a notion of stability with sibling complementarities and show how to compute stable matchings, while \citet{sun2025probabilistic} establish that stable matchings exist with high probability in large markets under similar priority structures. Relying on the stability concept developed by \citet{sun_stable_2024} and the existence justification provided by \citet{sun2025probabilistic}, this paper estimates household preferences in centralized assignment mechanisms with siblings.

Our paper also contributes to the literature on priority design in centralized assignment mechanisms. Theoretical work has largely treated priority as an exogenous input to the mechanism. One exception is \citet{celebi_priority_2022}, who study the optimal coarsening of priority levels when the underlying applicant attributes are continuous. Which priority structure is preferable depends on the distribution of applicant characteristics and is therefore inherently an empirical question, yet empirical evidence remains limited. \citet{kessel2018school} provide one of the few empirical studies by simulating alternative priority structures, including proximity-based rules, lottery-based rules, and reserved seats, using estimated student preferences in Swedish primary school choice and examining their effects on student welfare and segregation. Our contribution is to evaluate the welfare impact of a sibling-priority reform implemented in practice.

\section{Institutional Background}\label{sec:institution}

In Japan, access to licensed daycare is allocated through a publicly administered assignment system. Children aged 0 to 5 from households that meet eligibility requirements apply through their municipal government. Municipalities assign children to facilities using a centralized mechanism that considers each child's priority score and the household's reported preferences, subject to capacity constraints. Fees are largely uniform across licensed daycare facilities and are determined by household income rather than by individual facilities. Households may alternatively use nonlicensed daycare facilities, which are free to set prices and typically charge higher fees. For children aged 3 to 5, households may also choose kindergartens, which operate separately from licensed daycare and typically provide shorter daily hours, making them an imperfect substitute for full-day care.\footnote{Licensed childcare facilities include licensed nursery schools (\textit{hoikusho}), small-scale daycare facilities, workplace-based daycare facilities, and integrated early childhood education and care centers (\textit{kodomoen}). In this paper, kindergartens (\textit{yochien}) refer only to stand-alone kindergartens operating under the School Education Act and do not include integrated \textit{kodomoen}, which are part of the licensed daycare system.}

Koriyama City, the focus of this paper, is a mid-sized regional hub city located in Fukushima Prefecture in northeastern Japan. As of 2020, the city had a population of approximately 328,000 residents,\footnote{Koriyama City, \emph{Graph Koriyama Data Book 2025}, \url{https://www.city.koriyama.lg.jp/site/toukeikoriyama/5839.html}.} and private automobiles are the dominant mode of transportation, accounting for 82.8\% of commuting trips.\footnote{Statistics Bureau of Japan, \emph{2020 Population Census}, transportation-to-work-and-school results, \url{https://www.stat.go.jp/data/kokusei/2020/kekka.htm}.} There are about 12,000 children aged 0 to 5,\footnote{Koriyama City, \emph{Graph Koriyama Data Book 2025}, \url{https://www.city.koriyama.lg.jp/site/toukeikoriyama/5839.html}.} of whom roughly 1,400 apply each year to the centralized licensed daycare allocation mechanism.\footnote{Calculated from the administrative application data described in Section \ref{sec:data}.} As of 2025, the city had 90 licensed childcare facilities, 31 nonlicensed daycare facilities, and 32 kindergartens.\footnote{Calculated from the facility-level data described in Section \ref{sec:data}.}

In Koriyama City, households seeking daycare enrollment in April, which marks the start of the Japanese school year and therefore accounts for the majority of annual admissions, must submit applications in November of the preceding year. Before the application deadline, the municipality publishes information on coarse availability by age group for each licensed daycare facility. Each applicant then submits an application containing a preference report and the information from which the child's priority score is computed. The municipality then uses an assignment algorithm to allocate children subject to each facility's age-specific capacity constraints.

In the application form, a household reports a rank-ordered list (ROL) of up to ten preferred facilities for each child. Households applying for multiple children also specify two conditions:
(i) \textit{joint vs split assignment}: whether all placed children must attend the same facility;
(ii) \textit{full vs partial assignment}: whether the household accepts only all-or-none placement or permits some applying children to remain unassigned.
Together, these two conditions define the household's sibling-processing rule and create four possible preference patterns over sibling assignments.
For example, consider a household applying for two children. Suppose both children rank daycare $A$ first and $B$ second. 
Let $(x, y)$ denote the assignment pair in which the first child is assigned to daycare $x$ and the second to $y$, and let \(\varnothing\) denote no new placement. For each child, \(\varnothing\) means that the child retains their outside option, which need not be parental care.
Then, if the household requires both joint and full assignment, it prefers \((\varnothing,\varnothing)\) to \((A,B)\), \((B,A)\), \((A,\varnothing)\), \((B,\varnothing)\), \((\varnothing,A)\), and \((\varnothing,B)\).
If only joint assignment is required, then \((A,\varnothing)\), \((B,\varnothing)\), \((\varnothing,A)\), and \((\varnothing,B)\) are preferred to \((\varnothing,\varnothing)\).
If only full assignment is required, then \((A,B)\) and \((B,A)\) are preferred to \((\varnothing,\varnothing)\).
If neither is required, then \((A,B)\), \((B,A)\), \((A,\varnothing)\), \((B,\varnothing)\), \((\varnothing,A)\), and \((\varnothing,B)\) are preferred to \((\varnothing,\varnothing)\).
Together, the child-level ROLs and the household-level processing rule provide a partial ordering over assignment tuples.

Note that the message space of preference reporting in the application form introduced above is limited.
In the example above, a family that does not require joint assignment cannot express whether the family prefers \((A,B)\) to \((B,B)\).
The algorithm therefore imputes the remaining comparisons to construct a complete ordering; Appendix~\ref{sec:preference_completion} describes the imputation rule. It uses the number of children placed, whether they are placed together, and the children's reported facility rankings.
The application form also has a free-text section, in which the household is allowed to express more flexible preference patterns beyond what the individual ROLs and joint/split and full/partial preferences allow it to.
The Supplemental Appendix reproduces an English translation of the actual application form.

In the application form, the household also provides information used to compute each child's priority score, including parental employment status and sibling-related criteria. Importantly, ties in the total score are broken deterministically using additional criteria, such as working hours and single-parent status, which are considered lexicographically. We use these additional criteria to form the effective priority score defined in Section~\ref{sec:data}. The Supplemental Appendix presents a summary of the priority-scoring rules, including the tie-breaking rules.

These priority scores and preference reports are inputs to the municipal assignment algorithm. Since 2025, Koriyama City has used ChilmAI, provided by CyberAgent, Inc.; the detailed rules of the third-party algorithm used before 2025 are unavailable to us. For households applying for multiple children, ChilmAI converts the child-level ROLs and the sibling-processing rule into a complete ordering over assignment tuples, as mentioned above. Given age-specific capacities and household priority scores, it computes a matching stable with respect to these completed reports. Technically, it sets up a constrained programming problem to find a match, with constraints including the stability constraint. The relevant stability notion is that of \citet{sun_stable_2024}, which extends the usual notion of stability to a setting with siblings. We present its exact mathematical formulation in Section~\ref{sec:mechanism_stability}.

Households may have limited knowledge of how their reports affect assignments. Detailed priority-scoring rules are available on the municipal website, but they are posted separately and are not included in the applicant guide.\footnote{Koriyama City, ``Licensed Daycare Admission Information,'' \url{https://www.city.koriyama.lg.jp/site/kosodate/7048.html}.} The municipality only reports facility availability in three broad categories rather than as exact vacancy counts, and its applicant-facing materials do not explain in detail how the assignment algorithm uses a household's report. Given this information, a household may find it difficult to identify a profitable way to misreport its preferences.

\section{Data}\label{sec:data}

We use anonymized application data covering the years 2022 to 2025. The administrative data contain one record per applying child. For each child, we observe the parent ID, child ID, sibling status, the total priority score and its components (including maternal employment status, paternal employment status), the conditions used for tie-breaking, the reported rank-ordered list, the joint/split and full/partial assignment conditions reported by households applying for multiple children, the assigned daycare, and the household address.\footnote{Exact street numbers are masked. We use the centroid of the implied area as the household address.} In addition, the data record each child's current care status, such as parental care or enrollment in licensed or nonlicensed daycare or kindergarten. We interpret this status as the child's outside option, corresponding to the care arrangement that the child is assumed to use if the household ends up unassigned in the current application.

For each child, we construct an effective priority score by adding a fractional tie-breaking component to the integer priority score.
Within each year, we rank the tie-breaking values and normalize their ranks to obtain $\eta_c\in[0,1)$, with larger values indicating higher priority.
If child $c$'s integer priority score is $\tilde e_c$, the effective score used in the model is $e_c=\tilde e_c+\eta_c$.
This construction preserves the ordering of integer scores while incorporating the tie-breaking order.
In our structural analysis, we use this tie-broken, effective priority score.

Table \ref{tab:demographics_balance} presents summary statistics for applying children. We classify applicants into three groups based on sibling status: \textit{no-sibling} applicants, who apply for one child and have no enrolled siblings; \textit{simultaneous} applicants, who apply for multiple children in the same year; and \textit{incumbent} applicants, who apply while another child is already enrolled in a licensed daycare and is scheduled to remain enrolled in the following academic year. In 2025, 65.7\% of applying children are in the no-sibling group, 19.5\% in the simultaneous group, and 14.8\% in the incumbent group, implying that sibling households represent a substantial fraction of the applicant pool.\footnote{The descriptive tables retain all applying children. The 2025 household estimation and policy analysis exclude seven households applying for three children (21 child records), leaving 1,273 households.} While fathers' employment status shows little variation, with over 90\% employed full-time, mothers' employment status is more heterogeneous. In 2025, 77.8\% of mothers are employed full-time and 9.3\% part-time, with the remaining 13\% either searching for a job or not in the labor force. In 2025, nearly half of applicants (48.1\%) are in households with a parent on parental leave. Among these applicants, 99.4\% have mothers who are employed either full-time or part-time. Appendix \ref{sec:daycare_sumstats} reports the summary statistics for licensed childcare facilities.

\begin{table}[!htbp]
\centering
\caption{Applicant characteristics by year}
\label{tab:demographics_balance}
\footnotesize
\begin{threeparttable}
\begin{tabular}[t]{lcccc}
\toprule
Variable & 2022 & 2023 & 2024 & 2025\\
\midrule
Applying children & 1402 & 1442 & 1415 & 1422\\
~ ~ Age 0 & 381 & 365 & 368 & 296\\
~ ~ Age 1 & 550 & 581 & 559 & 609\\
~ ~ Age 2 & 141 & 165 & 175 & 173\\
~ ~ Age 3 & 231 & 235 & 217 & 259\\
~ ~ Age 4 & 62 & 63 & 63 & 64\\
~ ~ Age 5 & 37 & 33 & 33 & 21\\
Siblings &  &  &  & \\
~ ~ No-sibling & 0.631 & 0.645 & 0.634 & 0.657\\
~ ~ Simultaneous & 0.205 & 0.192 & 0.198 & 0.195\\
~ ~ Incumbent & 0.163 & 0.163 & 0.168 & 0.148\\
Mother's Employment Status &  &  &  & \\
~ ~ Full-Time & 0.688 & 0.739 & 0.767 & 0.778\\
~ ~ Part-Time & 0.149 & 0.135 & 0.111 & 0.093\\
~ ~ Seeking Employment & 0.128 & 0.099 & 0.085 & 0.101\\
~ ~ Not in Labor Force & 0.035 & 0.028 & 0.037 & 0.029\\
Father's Employment Status &  &  &  & \\
~ ~ Full-Time & 0.925 & 0.937 & 0.919 & 0.934\\
~ ~ Part-Time & 0.005 & 0.003 & 0.005 & 0.002\\
~ ~ Seeking Employment & 0.006 & 0.007 & 0.006 & 0.008\\
~ ~ Not in Labor Force & 0.064 & 0.053 & 0.069 & 0.056\\
Parental Leave & 0.437 & 0.436 & 0.499 & 0.481\\
\bottomrule
\end{tabular}
\begin{tablenotes}[flushleft]\footnotesize
\item Notes: Counts and shares use applying children as the unit of observation; household characteristics are repeated for each applying child. Age rows and Applying children are counts. All remaining rows report shares (means of indicators).
\end{tablenotes}
\end{threeparttable}
\end{table}

Table \ref{tab:app_summary_sibstat} reports summary statistics on application behavior by sibling status. Here, we focus on application behavior in 2025 to illustrate baseline patterns. Among simultaneous applicants, 54.2\% report joint/full assignment, 19.1\% joint/partial assignment, 10.5\% split/full assignment, and 16.2\% split/partial assignment. These shares use applying children as the unit of observation. Thus, 73.3\% require all placed children to attend the same facility, but some permit partial placement. Turning to incumbent applicants, 94.3\% list the daycare currently attended by the incumbent child as their top choice, and 63.0\% list only this daycare in their application. Taken together, these patterns provide clear evidence of a strong preference for joint assignment among applicants with siblings.

\begin{table}[p]
\centering
\caption{Summary by year and sibling status}
\label{tab:app_summary_sibstat}
\begin{threeparttable}
\centering
\fontsize{10}{12}\selectfont
\setlength{\tabcolsep}{4pt}
\begin{tabular*}{\textwidth}{@{\extracolsep{\fill}}lrrrrrr@{}}
\toprule
\multicolumn{7}{l}{\textit{Panel A: 2022 and 2023}} \\
\addlinespace[0.25em]
  & \multicolumn{3}{c}{2022} & \multicolumn{3}{c}{2023}
 \\
\cmidrule(l{3pt}r{3pt}){2-4}
\cmidrule(l{3pt}r{3pt}){5-7}
  & no-sibling & simultaneous & incumbent & no-sibling & simultaneous & incumbent
 \\
\midrule
Applying children & 885 & 288 & 229 & 930 & 277 & 235 \\
~~ Age 0 & 233 & 49 & 99 & 219 & 51 & 95 \\
~~ Age 1 & 366 & 79 & 105 & 401 & 66 & 114 \\
~~ Age 2 & 86 & 44 & 11 & 100 & 51 & 14 \\
~~ Age 3 & 150 & 69 & 12 & 148 & 76 & 11 \\
~~ Age 4 & 35 & 25 & 2 & 40 & 23 & - \\
~~ Age 5 & 15 & 22 & - & 22 & 10 & 1 \\
Score & 397.04 & 396.35 & 444.92 & 417.43 & 416.63 & 450.38 \\
Pref. Length & 3.20 & 3.48 & 1.88 & 3.32 & 2.99 & 1.83 \\
Joint / full & - & 0.66 & - & - & 0.53 & - \\
Joint / partial & - & 0.13 & - & - & 0.23 & - \\
Split / full & - & 0.07 & - & - & 0.10 & - \\
Split / partial & - & 0.14 & - & - & 0.09 & - \\
Assigned & 0.68 & 0.45 & 0.82 & 0.77 & 0.69 & 0.91 \\
Jointly Assigned & - & 0.39 & 0.73 & - & 0.62 & 0.81 \\
Assigned Rank & 1.58 & 1.99 & 1.11 & 1.51 & 1.66 & 1.04 \\
\bottomrule
\end{tabular*}
\vspace{0.8em}
\begin{tabular*}{\textwidth}{@{\extracolsep{\fill}}lrrrrrr@{}}
\toprule
\multicolumn{7}{l}{\textit{Panel B: 2024 and 2025}} \\
\addlinespace[0.25em]
  & \multicolumn{3}{c}{2024} & \multicolumn{3}{c}{2025}
 \\
\cmidrule(l{3pt}r{3pt}){2-4}
\cmidrule(l{3pt}r{3pt}){5-7}
  & no-sibling & simultaneous & incumbent & no-sibling & simultaneous & incumbent
 \\
\midrule
Applying children & 897 & 280 & 238 & 934 & 277 & 211 \\
~~ Age 0 & 215 & 51 & 102 & 180 & 49 & 67 \\
~~ Age 1 & 380 & 72 & 107 & 427 & 72 & 110 \\
~~ Age 2 & 102 & 52 & 21 & 107 & 48 & 18 \\
~~ Age 3 & 141 & 70 & 6 & 180 & 64 & 15 \\
~~ Age 4 & 36 & 25 & 2 & 32 & 31 & 1 \\
~~ Age 5 & 23 & 10 & - & 8 & 13 & - \\
Score & 446.01 & 606.77 & 585.30 & 448.45 & 612.48 & 599.32 \\
Pref. Length & 3.42 & 2.99 & 1.92 & 3.50 & 3.42 & 1.90 \\
Joint / full & - & 0.53 & - & - & 0.54 & - \\
Joint / partial & - & 0.20 & - & - & 0.19 & - \\
Split / full & - & 0.11 & - & - & 0.10 & - \\
Split / partial & - & 0.14 & - & - & 0.16 & - \\
Assigned & 0.71 & 0.69 & 0.86 & 0.78 & 0.87 & 0.91 \\
Jointly Assigned & - & 0.59 & 0.79 & - & 0.77 & 0.74 \\
Assigned Rank & 1.55 & 1.55 & 1.05 & 1.60 & 1.53 & 1.09 \\
\bottomrule
\end{tabular*}
\vspace{0.5em}
\parbox{\textwidth}{\footnotesize \textit{Notes:} The unit of observation is the applying child. Applying children and age rows report counts; Score, Pref. Length, and Assigned Rank report means; other rows report shares. The four reporting categories use all simultaneous-applicant child records as the denominator, including those with missing reports. Conditions are tabulated as recorded on each child's application. Missing reports affect 7 households (14 children) in 2023 and 3 households (5 children) in 2024. Joint requires all placed children to attend the same facility; split permits different facilities. Full permits only all-or-none placement; partial permits some but not all applying children to be placed. Jointly Assigned is an indicator equal to one if all children in the household are assigned to the same facility; its denominator includes all applying children in the sibling group, including those who are unassigned. Assigned Rank is the rank of the assigned facility in the reported rank-ordered list, conditional on being assigned. Cells with no observations are shown as ``-''.\par}
\end{threeparttable}
\end{table}

To further assess whether aversion to additional commuting distance contributes to the dislike of split assignment, we focus on simultaneous applicants and examine the geographical dispersion of daycare centers listed in an applicant's rank-ordered list, separately by whether the household requires joint assignment. The key idea is that households that do not require joint assignment face the possibility that their children may be placed at different centers and therefore must consider the total commuting distance across potential split placements. In contrast, households that require joint assignment do not permit their children to be placed at different centers and therefore do not face additional commuting distance arising from split placements. If commuting burden is an important driver of split assignment aversion, we would expect households that do not require joint assignment to list centers that are geographically closer to one another than households that require joint assignment.

\begin{figure}[htbp]
    \centering
    \includegraphics[width=0.68\linewidth]{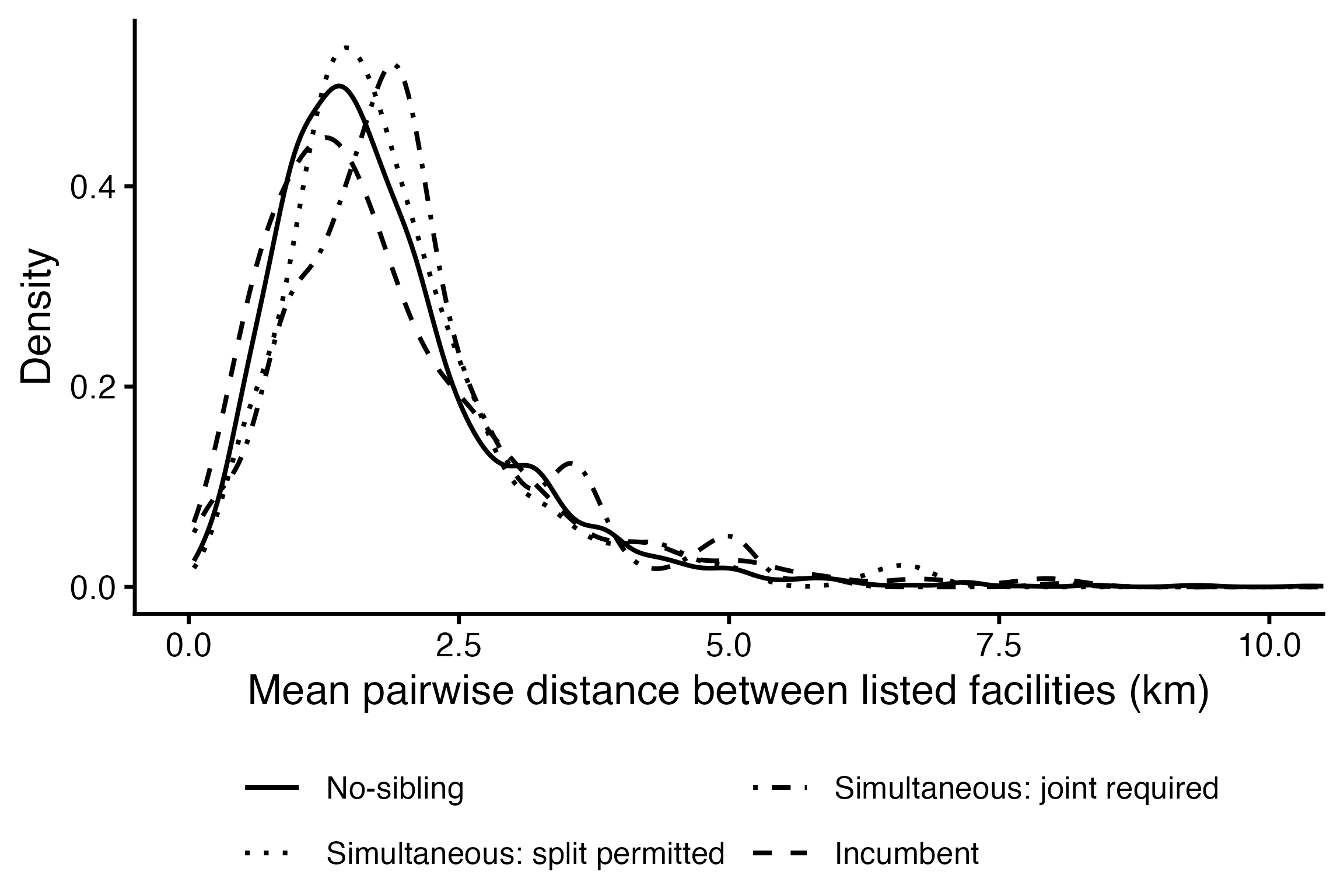}
    \caption{Geographic proximity of daycare choices by sibling status and joint assignment preference}
    \label{fig:rol_geo_concentration_density}
    \begin{minipage}{0.68\linewidth}
        \footnotesize
        \vspace{0.5em}
        \textit{Notes:} The figure plots kernel density estimates of the mean pairwise geographic distance between facilities listed in child-level rank-ordered lists in 2022--2025. Each list contains at least two geocoded facilities. All density curves are shown in black; line type distinguishes no-sibling applicants, simultaneous applicants who require joint assignment or permit split assignment, and incumbent applicants. Mean pairwise distance is computed using Haversine distances between geocoded facility locations. Lower values correspond to more geographically concentrated choice sets.
    \end{minipage}
\end{figure}

We measure the geographic proximity of daycare centers in an applicant's rank-ordered list using the mean pairwise Haversine distance (km) across all listed centers.\footnote{The Haversine formula computes great-circle distances between two points on the Earth's surface using their latitude and longitude coordinates, accounting for the curvature of the Earth. Distances are calculated in kilometers based on facility geocodes.} Figure \ref{fig:rol_geo_concentration_density} is consistent with the prediction derived above. It plots the distribution of this proximity measure by sibling status and stated joint assignment preference, and shows that applicants who do not require joint assignment tend to apply to more geographically concentrated sets of daycare centers. Table \ref{tab:qreg-geo-concentration} reports estimates from quantile regression models of the form \(Q_{\tau}(D \mid Joint) = \alpha_{\tau} + \beta_{\tau} Joint\), for \(\tau \in \{0.25, 0.50, 0.75\}\), where \(D\) denotes the proximity metric and \(Joint\) is an indicator for preferring joint assignment. The estimates indicate that, at the median, simultaneous applicants who prefer joint assignment list daycare centers that are approximately 0.198 km more geographically dispersed than those who do not, significant at the 10\% level. Taken together, these suggestive results motivate modeling applicant choice as reflecting not only individual trips but also the overall trip-chain distance associated with potentially multiple daycare placements.

\begin{table}[htbp]
\centering
\caption{Quantile regression of geographic dispersion on joint assignment preference}
\label{tab:qreg-geo-concentration}
\begin{threeparttable}
\newcommand{\sym}[1]{\ifmmode^{#1}\else\(^{#1}\)\fi}
\begin{tabular}{lccc}
\toprule
 & $\tau=0.25$ & $\tau=0.50$ & $\tau=0.75$ \\
\midrule
Joint & 0.077 & 0.198* & 0.180 \\
 & (0.122) & (0.107) & (0.194) \\
\addlinespace
Constant & 1.232*** & 1.660*** & 2.255*** \\
 & (0.086) & (0.097) & (0.161) \\
\midrule
Observations & \multicolumn{3}{c}{732} \\
\bottomrule
\end{tabular}
\begin{tablenotes}[flushleft]
\footnotesize
\item Notes: The 2022--2025 sample contains 732 child-level ROLs from 380 simultaneous-applicant household-years, each listing at least two geocoded facilities. The dependent variable is the mean pairwise Haversine distance (km) among daycare centers listed in the applicant's rank-ordered list. The regressor \textit{Joint} equals 1 if the applicant requires joint assignment and 0 otherwise. Each column reports a quantile regression at the indicated quantile. Bootstrap standard errors (1{,}000 replications), resampling household-years, are reported in parentheses. Significance: \sym{*} $p<0.10$, \sym{**} $p<0.05$, \sym{***} $p<0.01$.
\end{tablenotes}
\end{threeparttable}
\end{table}

\FloatBarrier

\section{Empirical Model}

Our empirical model of daycare choice extends standard school choice frameworks by allowing parents to value joint rather than individual assignments for their children. Following \citet{gentzkow_valuing_2007}, we model sibling complementarity in an additive manner relative to the sum of individual flow utilities. Family utility is specified as the sum of each child's flow utility minus a penalty when siblings are assigned to different daycare centers.

\subsection{Model Setup}

We index all children by \( c \in \overline{\mathcal{C}} \), families by \( f \in \mathcal{F} \), and facilities by \( d \in \mathcal{D} \). The set of facilities is partitioned as
\[
\mathcal{D} = \mathcal{D}^I \cup \mathcal{D}^O \cup \mathcal{D}^K \cup \{0\},
\]
where \( \mathcal{D}^I \) denotes licensed daycares, \( \mathcal{D}^O \) denotes nonlicensed daycares, \( \mathcal{D}^K \) denotes kindergartens, and \( d = 0 \) represents parental care at home.
Each child \( c \) belongs to a grade \( G(c) \in \mathcal{G} = \{0, 1, \dots, 5\} \).
For each licensed childcare facility \(d\) and grade \(g\), \(Q(d,g)\in\mathbb{N}_0\) denotes vacant seats available to new applicants.
We set \(Q(d,g)=\infty\) for an available nonlicensed daycare or kindergarten.\footnote{Some nonlicensed facilities and kindergartens serve only certain age groups. For an age group a facility does not serve, \(Q(d,g)=0\); the infinite-capacity convention applies to available nonlicensed facilities and kindergartens.}
Each child \( c \in \overline{\mathcal{C}} \) has an exogenous effective priority score \( e_c \in \mathbb{R} \), which is already tie-broken as described in Section~\ref{sec:data}, and a current assignment (including parental care) \( \overline{D}(c) \in \mathcal{D} \). Let \(\mathcal{C}\subseteq\overline{\mathcal{C}}\) denote the set of children for whom a family submits an application. 
The set of all children belonging to family \( f \) is denoted by \( \overline{C}(f) \subset \overline{\mathcal{C}} \), and the subset of applying children in family \( f \) is \(C(f)=\overline{C}(f)\cap \mathcal{C}\). We write \(e_f=(e_c)_{c\in C(f)}\) for the vector of applying children's priority scores.

The set of families can be partitioned into three mutually exclusive groups,
\[
\mathcal{F}
=
\mathcal{F}^{\text{no-sibling}}
\;\cup\;
\mathcal{F}^{\text{simultaneous}}
\;\cup\;
\mathcal{F}^{\text{incumbent}} .
\]
Here,
\[
\begin{aligned}
\mathcal{F}^{\text{no-sibling}}
&=
\{\, f \in \mathcal{F} \mid |\overline{C}(f)| = 1,\; |C(f)| = 1 \,\}, \\[0.5em]
\mathcal{F}^{\text{simultaneous}}
&=
\{\, f \in \mathcal{F} \mid |\overline{C}(f)| = 2,\; |C(f)| = 2 \,\}, \\[0.5em]
\mathcal{F}^{\text{incumbent}}
&=
\{\, f \in \mathcal{F} \mid |\overline{C}(f)| = 2,\; |C(f)| = 1 \,\}.
\end{aligned}
\]

We parameterize the family's preferences as follows.
For each child \(c \in \overline{C}(f)\) of family \(f\) in grade \(G(c)=g\), the flow utility from attending daycare \(d \in \mathcal{D}\setminus\{0\}\), before commuting cost, is
\[
u_{c,d}
=
\overline{\alpha}_g + \alpha_d + Z_f'\beta
+ X_d'\bigl(\Pi W_c + \Sigma\nu_f\bigr).
\]
Here, \(\overline{\alpha}_g\) is an age fixed effect and \(\alpha_d\) is a facility fixed effect. We normalize \(\overline{\alpha}_g=0\) for \(g=0\) and pool the age effects for ages four and five.\footnote{The age-five group contains only 19 applying children in the 2025 estimation sample, compared with 62 aged four.}
To reduce the number of facility parameters, we assume \(\alpha_d=\alpha_{d'}\) whenever \(d,d'\in\mathcal{D}^K\) or \(d,d'\in\mathcal{D}^O\). The vector \(Z_f\) contains indicators for the mother working full-time, the mother working part-time, and either parent being on parental leave. The two-element column vector \(X_d\) contains log total staff and log capacity, each standardized across licensed facilities. In our specification, \(W_c\) is an indicator that child \(c\) is aged two or younger, and \(\Pi\) is a two-element column vector containing \(\pi_{\mathrm{staff}}\) and \(\pi_{\mathrm{capacity}}\). These interactions allow preferences for facility staffing and capacity to differ for infants. The random coefficients allow households to differ in their preferences for these facility characteristics, with \(\Sigma=\operatorname{diag}(\sigma_{\mathrm{staff}},\sigma_{\mathrm{capacity}})\) and \(\nu_f\sim N(0,I_2)\). The same realization of \(\nu_f\) applies to all children in family \(f\). 
We normalize \(u_{c,0}=0\), so utilities are measured relative to parental care.

Given each child's flow utility, a family \(f\) has preferences over joint assignments for its children in
\(\mathcal{D}^{|\overline{C}(f)|}\).
For \(f \in \mathcal{F}^{\text{no-sibling}}\), the family's utility from assignment \(d\) equals the child's flow utility net of commuting cost plus an unobservable taste shock,
\[
u_{f,d} = u_{c,d} - \kappa\, d_{f,d} + \varepsilon_{f, d} .
\]
Here \(d_{f,d}\) denotes the distance in kilometers from family \(f\)'s residence to daycare \(d\).
Parental care involves no commuting, so \(d_{f,0} = 0\).
We assume that the shocks \(\varepsilon_{f, d}\) are independent and identically distributed according to a Type I extreme-value distribution with unit scale.

For \(f \in \mathcal{F}^{simultaneous} \cup \mathcal{F}^{incumbent}\), the utility of family \(f\) when child \(c\) is assigned to daycare \(d\) and child \(c'\) is assigned to daycare \(d'\) is
\[
u_{f,(d,d')}
=
u_{c,d} + u_{c',d'}
- \kappa\, d_{f,d,d'}
- \overline{\Gamma}_f \,\mathbb{I}\{d\neq d',\ d\neq0,\ d'\neq0\}
+ \varepsilon_{f, (d,d')} .
\]
The third term represents commuting distance. When both children attend childcare facilities, it is the shorter of the two possible trip chains. When one child remains in parental care, it is the distance from the family's residence to the other child's facility:
\[
d_{f,d,d'}
=
\begin{cases}
0, & d=d'=0,\\
d_{f,d'}, & d=0,\ d'\neq0,\\
d_{f,d}, & d\neq0,\ d'=0,\\
\min \bigl\{d_{f,d}+d_{d,d'},\ d_{f,d'}+d_{d',d}\bigr\},
& d\neq0,\ d'\neq0,
\end{cases}
\]
where \(d_{d,d'}\) denotes the distance in kilometers between daycare centers \(d\) and \(d'\).
For given \(f\) and \(d\), this distance is minimized when \(d' = d\), that is, when both children attend the same daycare and no additional commuting is required.
The fourth term captures the cost of assigning children to different facilities, beyond the extra commuting distance. It does not apply when one child remains in parental care. We parameterize this term as
\[
\overline{\Gamma}_f
=
\gamma_0
+ \gamma_{\mathrm{employed}}\mathbb{I}\{\text{mother employed}\}
+ \gamma_{\mathrm{leave}}\mathbb{I}\{\text{parental leave}\}.
\]
Here, mother employed means that the mother works either full-time or part-time.
We assume this cost is invariant across daycare pairs.
In the daycare context, parents are typically expected to prepare for a range of daycare-related activities, such as events, festivals, and the submission of daily diaries.
These requirements often differ across facilities in timing and format, generating a substantial coordination burden when siblings attend different daycare centers.
The shocks \(\varepsilon_{f,(d,d')}\) are independently and identically distributed according to a Type I extreme-value distribution with unit scale.

In general, the utility of family $f \in \mathcal{F}$ from being assigned to daycare tuple $\delta \in \mathcal{D}^{|\overline{C}(f)|}$ can be expressed as:
\begin{equation} \label{eq:decomposition_formula}
    u_{f,\delta}
    = V_{f,\delta} + \varepsilon_{f,\delta}
    = U_{f,\delta} - \Gamma_{f,\delta} - \kappa d_{f,\delta} + \varepsilon_{f,\delta}
\end{equation}
where \(U_{f,\delta}=\sum_{c\in\overline C(f)}u_{c,\delta(c)}\), and \(\delta(c)\) denotes the care arrangement assigned to child \(c\). The split-assignment term equals \(\overline{\Gamma}_f\) when children attend different facilities and zero otherwise:
\[
\Gamma_{f,\delta}
=
\overline{\Gamma}_f
\mathbb{I}
\left\{
\left|
\left\{
\delta(c):c\in\overline C(f),\ \delta(c)\neq0
\right\}
\right|
\geq2
\right\}.
\]

\subsection{Mechanism and Stability}\label{sec:mechanism_stability}

Initially, for each daycare $d$ and grade $g$, there is a number of vacant seats $Q(d,g)$. Prior to the application deadline, the municipality publicly announces information on expected availability.\footnote{As discussed in Section~\ref{sec:institution}, the municipality does not disclose the exact number of vacant seats. Instead, it reports a coarse availability indicator based on expected intake.} Each family $f$ submits a rank-ordered list of length $K_c$ for each child $c \in C(f)$, denoted by $L_c = (l_c^1, \dotsc, l_c^k, \dotsc, l_c^{K_c})$, where $l_c^k \in \mathcal{D}^{I}$ represents the daycare ranked $k$ in the child's ROL. Families applying for multiple children also submit a processing rule $m_f \in \{joint, split\} \times \{full, partial\}$. Specifically, we denote the joint-assignment component of this report by
\[
j_f
=
\mathbb I\{m_f \in \{ (joint, full), (joint, partial) \} \}.
\]

Given each family's rank-ordered lists $\{L_c\}_{c \in C(f)}$, sibling-processing rule $m_f$ for $f \in \mathcal{F}^{simultaneous}$, and priority score $e_c$, $c \in C(f)$, a matching determines the assignment of every child in the market, with non-applying children fixed at their current daycare. We denote this as a mapping
\[
\mu : \left( \{L_c, e_c\}_{c \in C(f)},\, m_f \right)_{f=1}^{|\mathcal F|}
\;\longrightarrow\;
\mathcal{D}^{|\overline{\mathcal{C}}|},
\]
where \(\mu(c)=\overline{D}(c)\) for all \(c \in \overline{\mathcal{C}} \setminus \mathcal{C}\).
An applying child who does not receive a new licensed placement retains \(\overline D(c)\).

The matching $\mu$ determines a cutoff score for each daycare $d$ and grade $g$, defined as follows:
\[
P_{d,g}(\mu) =
\begin{cases}
\min\{e_c \mid \mu(c) = d,\; G(c)=g,\; c \in \mathcal{C}\} & \text{if } \left|\mu^{-1}(d,g)\cap \mathcal{C}\right| = Q(d,g), \\[6pt]
0 & \text{if } \left|\mu^{-1}(d,g)\cap \mathcal{C}\right| < Q(d,g).
\end{cases}
\]

The market employs the algorithm proposed in \citet{sun_stable_2024}, which finds a stable matching based on the reported preferences, priority scores, and number of vacancies. We use their stability notion, stated below for our setting.
Consider a preference profile \((\succ_f)_{f \in \mathcal{F}}\), where each family \(f\) has strict preferences over tuples of daycare assignments in \(\mathcal{D}^{|\overline{C}(f)|}\).\footnote{Given the continuous distribution of the tuple shock, preferences are strict almost surely.}
\begin{samepage}
\begin{definition}[Stability]
Given a preference profile \((\succ_f)_{f \in \mathcal{F}}\), a feasible matching \(\mu\) is \textbf{stable} if and only if it admits no blocking coalitions. A family \(f\) forms a \textbf{blocking coalition} with a tuple of daycare centers \(\delta \in \mathcal{D}^{|\overline{C}(f)|}\) if and only if
(i) \(\delta \succ_f \mu(f)\),
(ii) \(\delta(c)=\overline{D}(c)\) for all \(c \in \overline{C}(f)\setminus C(f)\), and
(iii) \(e_c \ge P_{\delta(c), G(c)}(\mu)\) for all children \(c \in C(f)\).
\end{definition}
\end{samepage}
The cutoff condition is applied separately to each child.\footnote{The individual cutoff condition may not ensure joint feasibility when two siblings of the same age seek the same facility. This issue can arise for at most 13 households: the 13 two-child households whose children are in the same age group, representing 10.2\% of the 128 simultaneous-applicant households and 1.0\% of all 1,273 households.}
The algorithm uses constraint programming to compute a matching that is stable with respect to reported preferences. Because the mechanism restricts each applicant's message space to individual ROLs and one of four sibling-processing rules, it completes the household's partial ordering over daycare tuples using the rule documented in Appendix~\ref{sec:preference_completion}.

\begin{assumption}[Stability under latent preferences]\label{ass:latent_stability}
The observed matching \(\mu\) is stable with respect to families' latent preferences, as represented by \(u_{f,\delta}\).
\end{assumption}

This assumption entails both that a stable matching exists under latent preferences and that the observed matching is one such matching. On existence, \citet{sun2025probabilistic} show that stability is highly likely in large markets when facilities have similar priorities over children. Priorities in our setting are uniform across facilities, and the market contains more than one thousand applicant households. The second part is not guaranteed by the algorithm, which uses submitted rather than latent preferences. However, a facility's position in a child's ROL does not affect the child's priority, while applicants have only limited information about market competitiveness. These features make profitable misreports difficult to identify. We maintain that differences between submitted and latent preferences do not create a blocking tuple at the observed matching.

\subsection{Likelihood Function}
Let $\mathcal{A}_{f}(e_f, \mu) \subseteq \mathcal{D}^{|\overline{C}(f)|}$ denote the collection of tuples $\delta$ of daycares such that $\delta(c)=\overline{D}(c)$ for all $c \in \overline{C}(f)\setminus C(f)$ and $e_c \geq P_{\delta(c),G(c)}(\mu)$ for all $c \in C(f)$. Then an implication of a stable matching $\mu$ is that family $f$ chooses the tuple that maximizes its utility among the elements of $\mathcal{A}_f(e_f,\mu)$:
\[
\mu(f)=\argmax_{\delta \in \mathcal{A}_f(e_f,\mu)} u_{f,\delta}.
\]
This allows us to reformulate the problem as a discrete choice problem with personalized choice sets, as in \citet{fack_beyond_2019}.

Let \(\tilde Z_f\) denote the vector of observed covariates for family \(f\), including \(Z_f\) and the ages \(G(c)\) for all \(c \in \overline{C}(f)\). Let \(\vartheta\) collect the preference parameters. Because the systematic utility \(V_{f,\delta}(\nu_f;\vartheta)\) depends on the household random coefficient \(\nu_f\), we first write the choice probability conditional on its realization. Given that the tuple-level taste shocks are independently and identically distributed according to a Type I extreme-value distribution, this conditional probability is
\begin{align*}
&\mathbb{P}\bigl(\delta = \mu(f) \mid \nu_f, \tilde Z_f, e_f, \mathcal{A}_f(e_f,\mu); \vartheta\bigr) \\
&\quad= \mathbb{P}\!\left(
\delta = \arg\max_{\delta' \in \mathcal{A}_f(e_f,\mu)} u_{f,\delta'}
\;\middle|\;
\nu_f, \tilde Z_f, e_f, \mathcal{A}_f(e_f,\mu); \vartheta
\right) \\
&\quad= \frac{\exp\!\bigl(V_{f,\mu(f)}(\nu_f;\vartheta)\bigr)}
{\sum_{\delta' \in \mathcal{A}_f(e_f,\mu)} \exp\!\bigl(V_{f,\delta'}(\nu_f;\vartheta)\bigr)}.
\end{align*}
We assume that, conditional on \(\tilde Z_f\), the priority score and personalized choice set are independent of both sources of unobserved preference heterogeneity:
\[
\bigl(e_f,\mathcal{A}_f(e_f,\mu)\bigr)
\;\perp\!\!\!\perp\;
\bigl(\nu_f,\{\varepsilon_{f,\delta}\}_{\delta}\bigr)
\;\mid\; \tilde Z_f.
\]
The priority score is based on observable characteristics, such as parental employment status. For a single-child applicant, a common priority order across daycares means that the available facilities are determined by higher-priority applicants' choices, not by the applicant's own unobserved preferences \citep{fack_beyond_2019}. We maintain the analogous conditional independence of the personalized choice set for households applying with multiple children.

Integrating over the standard bivariate normal density \(\phi(\nu)\), the unconditional probability of the observed assignment is
\begin{equation}
\label{eq:assignment_probability}
P_f\bigl(\mu(f);\vartheta\bigr)
=
\int_{\mathbb{R}^2}
\frac{\exp\!\bigl(V_{f,\mu(f)}(\nu;\vartheta)\bigr)}
{\sum_{\delta' \in \mathcal{A}_f(e_f,\mu)}
\exp\!\bigl(V_{f,\delta'}(\nu;\vartheta)\bigr)}
\phi(\nu)\,d\nu.
\end{equation}
Our likelihood function is therefore
\[
L\bigl(\vartheta \mid
\{\tilde Z_f,e_f,\mathcal{A}_f(e_f,\mu)\}_{f\in\mathcal F}\bigr)
=
\prod_{f\in\mathcal F}P_f\bigl(\mu(f);\vartheta\bigr),
\]
and the corresponding log likelihood is
\[
\ell(\vartheta)
=
\sum_{f\in\mathcal F}
\log P_f\bigl(\mu(f);\vartheta\bigr).
\]

\section{Estimation}

The likelihood representation implied by stability allows us to learn household preferences by treating observed assignments as choices from household-specific feasible sets. Assignment data alone, however, may provide limited information for distinguishing the disutility from split assignment from household-level heterogeneity in facility preferences. For example, observing two siblings assigned to the same facility can be rationalized either by a large split-assignment cost or by a strong shared preference for that facility arising from the household random coefficients. The submitted ROLs and sibling-processing rules provide additional information that helps distinguish these explanations. We therefore follow the moment-based approach of \citet{fack_beyond_2019}: we express the information in the stability-based likelihood as score equalities and combine them with moment inequalities derived from reported preferences.

\subsection{Converting Likelihood into Score Moments}
Let \(\vartheta_k\) denote the \(k\)th element of the parameter vector \(\vartheta\), for \(k=1,\ldots,K\). Differentiating the logarithm of family \(f\)'s assignment probability in equation~\eqref{eq:assignment_probability} with respect to \(\vartheta_k\) gives its score contribution,
\[
s_{fk}(\vartheta)
=
\frac{\partial \log P_f\bigl(\mu(f);\vartheta\bigr)}
{\partial \vartheta_k}.
\]
Under standard regularity conditions, the score has mean zero at the true parameter vector \(\vartheta_0\):
\[
E\!\left[s_{fk}(\vartheta_0)\right]=0,
\qquad k=1,\ldots,K.
\]
The corresponding sample moment is
\[
\overline{s}_k(\vartheta)
=
\frac{1}{|\mathcal F|}
\sum_{f\in\mathcal F}s_{fk}(\vartheta),
\qquad k=1,\ldots,K.
\]

\subsection{Bounds from the Sibling-Processing Rule}
\label{sec:processing_rule_bounds}

The following assumption relates the reported joint-assignment requirement to the household's utility comparison between split assignment and joint non-assignment. With a slight abuse of notation, write \(u_{f,(\varnothing,\varnothing)}\equiv u_{f,(\overline D(c),\overline D(c'))}\) for utility when neither child receives a new placement and each retains their outside option.

\begin{samepage}
\begin{assumption}[Sibling-processing-rule preservation]
\label{ass:processing_rule_preservation}
For two children $c$ and $c'$ and facilities $d\neq d'$ such that $d\in L_c$ and $d,d'\in L_{c'}$,
\[
\begin{aligned}
j_f=1 &\quad\Longrightarrow\quad u_{f,(\varnothing,\varnothing)}\geq u_{f,(d,d')},\\
j_f=0 &\quad\Longrightarrow\quad u_{f,(d,d')}\geq u_{f,(\varnothing,\varnothing)}.
\end{aligned}
\]
\end{assumption}
\end{samepage}

The restriction says that, for facilities satisfying the ROL condition, a household requesting joint assignment prefers joint non-assignment to every split assignment, while a household permitting split assignment prefers every split assignment to joint non-assignment. It can fail in either direction. A household requesting joint assignment may nevertheless accept split assignment at a particular pair of facilities, while a household permitting split assignment may regard split assignment at a particular pair of facilities as worse than joint non-assignment. The utility model does not rule out either possibility.
As discussed in Section~\ref{sec:institution}, households can state such a flexible ordering in the application's free-text field. Based on information provided by municipal staff, we confirmed that such requests occurred in only two of the 128 simultaneous-applicant households in our 2025 estimation sample.\footnote{We retain these two households in the processing-rule moments and code their structured reports in the same way as other households.}

We use the reported sibling-processing rule to bound the probability that joint non-assignment is preferred to a particular split-assignment tuple.
Throughout this subsection, probabilities are conditional on observed household and facility characteristics; we suppress this conditioning to keep the notation concise.
Define
\[
p_{\Gamma,f}^{dd'}(\vartheta)
=
\Pr_{\vartheta}\!\left(
u_{f,(\varnothing,\varnothing)}>u_{f,(d,d')}
\right).
\]
Assumption~\ref{ass:processing_rule_preservation} implies
\[
\Pr\!\left(
\substack{
j_f=1,\\
d\in L_c,\ d,d'\in L_{c'}
}
\right)
\leq
p_{\Gamma,f}^{dd'}(\vartheta)
\leq
1-
\Pr\!\left(
\substack{
j_f=0,\\
d\in L_c,\ d,d'\in L_{c'}
}
\right).
\]
The lower (upper) bound follows because $j_f=1$ ($j_f=0$) implies that utility from $(d,d')$ is below (above) $(\varnothing,\varnothing)$.

For any nonnegative instrument vector $h_f^{dd'}$, the corresponding unconditional moment inequalities hold componentwise:
\[
\E\!\left[
h_f^{dd'}
\left\{
p_{\Gamma,f}^{dd'}(\vartheta)
-
\mathbb I\!\left\{
\substack{
j_f=1,\ d\in L_c,\\
d,d'\in L_{c'}
}
\right\}
\right\}
\right]
\geq\boldsymbol 0,
\]
and
\[
\E\!\left[
h_f^{dd'}
\left\{
1-
\mathbb I\!\left\{
\substack{
j_f=0,\ d\in L_c,\\
d,d'\in L_{c'}
}
\right\}
-
p_{\Gamma,f}^{dd'}(\vartheta)
\right\}
\right]
\geq\boldsymbol 0.
\]
Note that the report indicator is informative only when the assignment tuple $(d,d')$ satisfies the ROL condition $d\in L_c$ and $d,d'\in L_{c'}$.\footnote{In two other households, the children's records disagree on both the full/partial and joint/split conditions. This is a conflict between recorded household-level entries. We exclude these households only from the processing-rule inequalities; they remain in the assignment-likelihood score moments, the sibling-ROL inequalities in Section~\ref{sec:rol_bounds}, and welfare analysis.}

With $J=90$ licensed facilities, 4,005 facility pairs are possible, but many are too rarely listed by both children to inform a pair-specific bound. We retain a pair if at least five households list both facilities on both children's ROLs; 30 pairs meet this requirement. This leads us to consider both child-specific assignments for each pair, yielding a set $\Delta$ of 60 ordered tuples. For each $(d,d')\in\Delta$, we form a local moment using households living within two kilometers of both facilities. This overlapping catchment focuses the comparison on households for whom the pair is geographically relevant. The 66-component instrument vector $h_f^{dd'}$ has 60 pair-specific entries: only the entry for $(d,d')$ can be nonzero, and it equals the catchment indicator. Its remaining six entries equal that indicator times a constant, the youngest and oldest children's ages divided by five, indicators for maternal part-time and full-time employment, and parental-leave status, each divided by the number of retained ordered tuples in the household's catchment. These six entries average contributions across nearby tuples within each household.

Replacing the population expectations with household averages, the vector of lower-bound sample moments is
\begin{equation}
\label{eq:sample_report_moment}
\begin{aligned}
\overline{\boldsymbol m}_{\mathrm{rule,lower}}(\vartheta)
&=
\frac{1}{|\mathcal F^{\text{simultaneous}}|}
\sum_{f\in\mathcal F^{\text{simultaneous}}}
\sum_{(d,d')\in\Delta}
h_f^{dd'}\\
&\quad\times
\left[
p_{\Gamma,f}^{dd'}(\vartheta)
-\mathbb I\!\left\{
j_f=1,\ d\in L_c,\ d,d'\in L_{c'}
\right\}
\right]
\geq\boldsymbol 0.
\end{aligned}
\end{equation}
Replacing the bracket by $1-\mathbb I\{j_f=0,\ d\in L_c,\ d,d'\in L_{c'}\}-p_{\Gamma,f}^{dd'}(\vartheta)$ defines the upper-bound vector $\overline{\boldsymbol m}_{\mathrm{rule,upper}}(\vartheta)$. Let $\mathcal M_{\mathrm{rule}}$ index the $2(60+6)=132$ scalar entries of these two vectors.

\subsection{Bounds from Sibling ROLs}\label{sec:rol_bounds}

We next use the similarity of the two ROLs to discipline shared household taste heterogeneity. Stronger shared tastes make the two children more likely to have similar utility orderings over facilities. To bound this probability, we require the submitted order for each child to preserve the corresponding child-specific utility comparison.

\begin{assumption}[Child-level ROL-order preservation]
\label{ass:rol_order_preservation}
For any facilities $d$ and $d'$ appearing on the same child's ROL,
\[
d \succ_{L_c} d'
\quad\Longrightarrow\quad
u_{c,d}-\kappa d_{f,d}
\geq
u_{c,d'}-\kappa d_{f,d'}.
\]
\end{assumption}

Here, $d\succ_{L_c}d'$ means that child $c$'s submitted ROL ranks $d$ above $d'$. Assumption~\ref{ass:rol_order_preservation} does not assert that households report individual ROLs truthfully. Rather, it requires only that a household not reverse the child-specific utility ordering when writing the child's ROL. The order-imputation rule described in Appendix~\ref{sec:preference_completion} can make some reversals of a child's reported ranking irrelevant to the household's assignment. For example, suppose that the older child's ROL is $(A,B)$ and the younger child's ROL is $(B,A)$. Changing the younger child's report from $(B,A)$ to $(A,B)$ does not change the ordering among assignment tuples in which both children are assigned, whether or not split assignment is permitted.

Under Assumption~\ref{ass:rol_order_preservation}, the reported ROLs bound the probability that both children prefer $d$ to $d'$. Define
\[
p_f^{dd'}(\vartheta)
=
\Pr_{\vartheta}\!\left(
u_{c,d}-\kappa d_{f,d}
\geq
u_{c,d'}-\kappa d_{f,d'},\;
u_{c',d}-\kappa d_{f,d}
\geq
u_{c',d'}-\kappa d_{f,d'}
\right).
\]
The observable reports imply
\[
\Pr\!\left(
d \succ_{L_c} d',\;
d \succ_{L_{c'}} d'
\right)
\leq
p_f^{dd'}(\vartheta)
\leq
1-
\Pr\!\left(
d' \succ_{L_c} d
\text{ or }
d' \succ_{L_{c'}} d
\right).
\]
The lower bound follows because agreement in the two reported rankings implies agreement in the two latent utility comparisons. The upper bound follows because a report of $d'$ above $d$ by either child rules out both children preferring $d$ to $d'$. The lower-bound indicator requires both facilities on both children's ROLs; the upper-bound indicator is informative if either child lists both facilities and ranks $d'$ above $d$.

Using the instruments introduced in Section~\ref{sec:processing_rule_bounds}, these probability bounds yield the unconditional moment inequalities
\[
\E\!\left[
h_f^{dd'}
\left\{
p_f^{dd'}(\vartheta)
-
\mathbb I\!\left\{
d \succ_{L_c} d',\;
d \succ_{L_{c'}} d'
\right\}
\right\}
\right]
\geq\boldsymbol 0,
\]
and
\[
\E\!\left[
h_f^{dd'}
\left\{
1
-
\mathbb I\!\left\{
d' \succ_{L_c} d
\text{ or }
d' \succ_{L_{c'}} d
\right\}
-p_f^{dd'}(\vartheta)
\right\}
\right]
\geq\boldsymbol 0.
\]

The corresponding sample vectors $\overline{\boldsymbol m}_{\mathrm{ROL,lower}}(\vartheta)$ and $\overline{\boldsymbol m}_{\mathrm{ROL,upper}}(\vartheta)$ are defined analogously to equation~\eqref{eq:sample_report_moment}, using the lower- and upper-bound expressions above and averaging over simultaneous-applicant households. Let $\mathcal M_{\mathrm{ROL}}$ index their 132 scalar entries.

\subsection{Objective Function and Inference}

Let $\overline m_{\ell}(\vartheta)$ denote a scalar entry of the sample inequality-moment vectors constructed above, and let $\mathcal M$ contain the indices of the entries included in the specification. Let $\widehat\omega_{s,k}$ and $\widehat\omega_{m,\ell}$ denote fixed normalization factors for the score and inequality moments. As in \citet{andrews_shi_2013} and \citet{fack_beyond_2019}, we minimize
\[
\widehat Q(\vartheta)
=
\sum_{k=1}^{K}
\left(
\frac{\overline s_k(\vartheta)}{\widehat\omega_{s,k}}
\right)^2
+
\sum_{\ell\in\mathcal M}
\left[
\min\!\left\{
\frac{\overline m_{\ell}(\vartheta)}{\widehat\omega_{m,\ell}},0
\right\}
\right]^2.
\]
We compute $\widehat\omega_{s,k}$ and $\widehat\omega_{m,\ell}$ as the standard deviations of the corresponding household-level moment contributions. These are evaluated at an initial parameter estimate obtained by maximizing the assignment likelihood without the reported-preference inequalities, and held fixed when minimizing the objective function.\footnote{We use 256 sign-balanced normal draws per household (per child in the individual-choice model), minimize using box-constrained L-BFGS from alternative starting values, and retain estimates satisfying a projected-gradient tolerance of $10^{-3}$.}

We report local sandwich standard errors. For the first three specifications, let $\overline g_A(\vartheta)$ stack the standardized score moments and the inequalities that are violated at the point estimate. Let $\widehat G_A$ denote its Jacobian and $\widehat\Omega_A$ the household-cluster covariance matrix of the corresponding moment contributions. Then
\begin{equation}
\label{eq:sandwich_covariance}
\widehat{\operatorname{Var}}(\widehat\vartheta)
=
\frac{1}{|\mathcal F|}
(\widehat G_A'\widehat G_A)^{-1}
\widehat G_A'\widehat\Omega_A\widehat G_A
(\widehat G_A'\widehat G_A)^{-1}.
\end{equation}
This calculation holds the estimated active inequality set fixed.

The baseline specification includes the score moments and uses $\mathcal M=\mathcal M_{\mathrm{rule}}\cup\mathcal M_{\mathrm{ROL}}$. The two alternatives also include the score moments but use $\mathcal M=\mathcal M_{\mathrm{rule}}$ and $\mathcal M=\mathcal M_{\mathrm{ROL}}$, respectively. We also estimate a canonical individual-choice model that treats each applying child as a separate single-child household, with independent random-coefficient draws across children. This benchmark is estimated using the assignment likelihood alone.\footnote{The likelihood sandwich clusters child scores by original household.}

\section{Results}\label{sec:results}

Table \ref{tab:estimates} reports the estimation results. Point estimates are expressed relative to the unit scale of the tuple shock $\epsilon$. Column 1 reports our baseline specification, which uses both the sibling-processing-rule bounds and the ROL-similarity bounds. Columns 2 and 3 use only the sibling-processing-rule bounds and only the ROL-similarity bounds, respectively. Column 4 reports the individual-choice model. Estimates of daycare fixed effects are reported in Appendix \ref{sec:alphad_estimates}. Parentheses report household-clustered analytical standard errors computed using equation~\eqref{eq:sandwich_covariance}.

At the reported estimates, one of the 264 inequality moments contributes a nonzero penalty in the specification using both restrictions. The specification using only the sibling-processing-rule bounds has one such moment out of 132, while the specification using only the ROL-similarity bounds has none out of 132. Thus, most inequality restrictions are satisfied at the point estimates. The reported standard errors hold the estimated active set fixed and provide a local approximation that does not account for changes in that set across samples.

\begin{table}[!htbp]
\centering
\caption{Structural preference estimates}
\label{tab:estimates}
\scriptsize
\setlength{\tabcolsep}{4pt}
\begin{threeparttable}
\begin{tabular}{lcccc}
\toprule
Parameter & \makecell{Both report\\bounds} & \makecell{Sibling-processing-rule\\bounds} & \makecell{ROL-similarity\\bounds} & \makecell{Individual-choice\\model} \\
 & (1) & (2) & (3) & (4) \\
\midrule
\multicolumn{5}{l}{\textbf{Age fixed effects ($\overline{\alpha}_g$)}} \\
~~Age 0 & \makecell{0.000 \\ (--)} & \makecell{0.000 \\ (--)} & \makecell{0.000 \\ (--)} & \makecell{0.000 \\ (--)} \\
~~Age 1 & \makecell{-3.758 \\ (1.421)} & \makecell{-2.928 \\ (0.900)} & \makecell{-3.529 \\ (1.074)} & \makecell{-2.023 \\ (0.617)} \\
~~Age 2 & \makecell{-2.026 \\ (1.541)} & \makecell{-1.391 \\ (1.131)} & \makecell{-1.792 \\ (1.200)} & \makecell{-0.714 \\ (0.764)} \\
~~Age 3 & \makecell{-1.536 \\ (1.500)} & \makecell{-0.708 \\ (1.009)} & \makecell{-1.353 \\ (1.267)} & \makecell{-0.058 \\ (0.732)} \\
~~Ages 4--5 & \makecell{-1.129 \\ (1.705)} & \makecell{-0.225 \\ (1.314)} & \makecell{-0.949 \\ (1.434)} & \makecell{0.595 \\ (1.033)} \\
\midrule
\multicolumn{5}{l}{\textbf{Flow utility covariates ($\beta$)}} \\
~~Mother part-time & \makecell{1.787 \\ (0.623)} & \makecell{1.813 \\ (0.593)} & \makecell{1.804 \\ (0.606)} & \makecell{1.845 \\ (0.584)} \\
~~Mother full-time & \makecell{2.717 \\ (0.638)} & \makecell{2.583 \\ (0.522)} & \makecell{2.704 \\ (0.483)} & \makecell{2.438 \\ (0.420)} \\
~~Parental leave & \makecell{-2.245 \\ (0.435)} & \makecell{-2.148 \\ (0.388)} & \makecell{-2.214 \\ (0.384)} & \makecell{-2.185 \\ (0.357)} \\
~~Infant $\times$ staff & \makecell{-0.610 \\ (1.064)} & \makecell{-0.973 \\ (1.073)} & \makecell{-0.641 \\ (0.344)} & \makecell{-0.283 \\ (0.130)} \\
~~Infant $\times$ capacity & \makecell{-0.010 \\ (0.492)} & \makecell{0.801 \\ (0.272)} & \makecell{-0.120 \\ (0.424)} & \makecell{0.703 \\ (0.230)} \\
\midrule
\multicolumn{5}{l}{\textbf{Random-coefficient heterogeneity}} \\
SD: staff coefficient & \makecell{3.492 \\ (5.876)} & \makecell{3.979 \\ (5.002)} & \makecell{3.395 \\ (0.843)} & \makecell{0.001$^{\dagger}$ \\ (--)} \\
SD: capacity coefficient & \makecell{2.583 \\ (0.630)} & \makecell{0.061 \\ (0.993)} & \makecell{2.433 \\ (0.550)} & \makecell{0.001$^{\dagger}$ \\ (--)} \\
\midrule
\multicolumn{5}{l}{\textbf{Split-assignment cost ($\gamma$)}} \\
~~Intercept & \makecell{4.526 \\ (1.279)} & \makecell{4.332 \\ (0.890)} & \makecell{4.561 \\ (0.989)} & -- \\
~~$\times$ mother employed & \makecell{-0.598 \\ (1.358)} & \makecell{-0.512 \\ (1.037)} & \makecell{-0.659 \\ (1.070)} & -- \\
~~$\times$ parental leave & \makecell{0.790 \\ (0.870)} & \makecell{0.622 \\ (0.782)} & \makecell{0.779 \\ (0.586)} & -- \\
\midrule
\multicolumn{5}{l}{\textbf{Other parameters}} \\
Distance penalty ($\kappa$) & \makecell{0.981 \\ (0.114)} & \makecell{0.923 \\ (0.089)} & \makecell{0.976 \\ (0.088)} & \makecell{0.817 \\ (0.052)} \\
\bottomrule
\end{tabular}
\begin{tablenotes}[flushleft]
\footnotesize
\item Notes: Columns 1--3 report the household model using, respectively, both sets of report bounds, only the sibling-processing-rule bounds, and only the ROL-similarity bounds. Column 4 treats each child as a separate decision maker. Household-clustered analytical standard errors are in parentheses. For Columns 1--3, the local sandwich calculation conditions on the estimated active inequality set. A dash denotes a normalized or inapplicable coefficient, or a standard error not reported for a boundary estimate. $\dagger$ marks an estimate at the numerical lower bound of 0.001 for the standard deviation; no standard error is reported. Daycare fixed effects are omitted.
\end{tablenotes}
\end{threeparttable}
\end{table}

We first focus on the baseline model in Column 1. The estimated coefficient on commuting distance, $\kappa$, is 0.981 and highly statistically significant. Because distance enters utility with a negative sign, an additional kilometer of commuting lowers household utility by 0.981 utility units. This estimate implies a strong preference for daycare centers located closer to home.

The parameters $\beta$, which govern the average utility of childcare services, can be interpreted as reflecting the opportunity cost of home-based parental care across household types. Relative to a household in which the mother is not employed, the utility of childcare is 2.717 units higher when the mother works full-time and 1.787 units higher when she works part-time. Both estimates are statistically significant. The coefficient on parental leave is $-2.245$ and statistically significant, consistent with parental care being less costly while a parent is temporarily away from work. By contrast, the interactions of infant status (ages 0--2) with facility staff and capacity are small relative to their standard errors and statistically insignificant. We therefore do not find strong evidence that average preferences for these facility characteristics differ between infants and older children.

We do find evidence of unobserved household heterogeneity in facility preferences, although its precision differs across characteristics. The estimated standard deviation of the capacity coefficient is 2.583, with a standard error of 0.630. The estimated standard deviation of the staff coefficient is also large, at 3.492, but its standard error is 5.876. Thus, the baseline specification provides strong evidence of heterogeneity in preferences for facility capacity, while the extent of heterogeneity in preferences for staffing is imprecisely estimated.

Turning to the complementarity parameters $\gamma$, the estimated constant term is large and positive, at 4.526, with a standard error of 1.279. Normalizing this estimate by the distance coefficient $\kappa$ gives a distance equivalent of $4.526/0.981=4.61$ kilometers for the reference household. In our sample, the mean modeled commuting distance among households with positive distance is 2.27 kilometers, so the point estimate of the non-distance penalty from split assignment is about twice this mean. Assigning siblings to different daycare centers therefore generates a utility loss substantially larger than a typical commute, over and above the direct increase in travel distance itself. The interactions of the split-assignment cost with maternal employment and parental leave are $-0.598$ and 0.790, respectively, and neither is statistically significant. We therefore do not find strong evidence of observed heterogeneity in the fixed cost of split assignment.

The main estimates are remarkably stable across the first three specifications. The coefficients on maternal employment and parental leave change little. The distance penalty ranges from 0.923 to 0.981, and the baseline split-assignment cost ranges from 4.332 to 4.561. The random coefficients show more clearly what the ROL-similarity bounds contribute. With only the sibling-processing-rule bounds, both random-coefficient standard deviations are imprecisely estimated, and the capacity standard deviation is close to zero. With only the ROL-similarity bounds, the staff and capacity standard deviations are 3.395 and 2.433, with standard errors of 0.843 and 0.550. When both sets of bounds are imposed, the capacity heterogeneity remains large and precisely estimated, although the staff component becomes imprecise. This pattern is consistent with the ROL-similarity bounds providing information about the magnitude of shared household heterogeneity in facility preferences.

Finally, the individual-choice model in Column 4 estimates a smaller distance penalty of 0.817. Households with siblings may accept more distant daycare centers to avoid split assignment when only one child can be admitted nearby. A model that treats each child as an independent decision maker attributes part of this coordination-driven behavior to weaker distance sensitivity. The lower distance penalty in the individual-choice model is consistent with this interpretation.

\section{Policy Evaluation}\label{sec:counterfactual}

\subsection{The 2024 Sibling Priority Reform}
The estimates in Section \ref{sec:results} indicate substantial disutility from assigning siblings to different facilities. Before the reform, Koriyama City granted no sibling-priority points to simultaneous applicants and only 25 points to incumbent applicants. In 2023, the unconditional joint-assignment rates were 62\% for simultaneous applicants and 81\% for incumbent applicants (Table \ref{tab:app_summary_sibstat}). Koriyama City viewed the difficulty of joint assignment, especially among simultaneous households, as a policy concern and consulted us on the appropriate amount of priority to grant households with siblings.
We simulated assignments under a menu of alternative uniform sibling-priority point combinations for simultaneous and incumbent households. From this menu, the city selected 160 points for both groups. The final scheme instead set the simultaneous-applicant bonus at 200 points when at least two siblings were of the same age and incorporated several other adjustments, documented in the Supplemental Appendix.
The reform applied to applications for admission beginning in 2024. Koriyama City announced the reform on its website as a revision intended to make it easier for siblings to attend the same daycare facility.\footnote{Koriyama City's announcement, ``Revising the criteria to make it easier for siblings to attend the same daycare facility,'' November 8, 2023. See \url{https://www.city.koriyama.lg.jp/site/kosodate/92957.html}.} For details on the simulation design and practical implementation, see \citet{Moriwaki2025Childcare}.

\begin{figure}[htbp]
    \centering
    \includegraphics[width=0.82\linewidth]{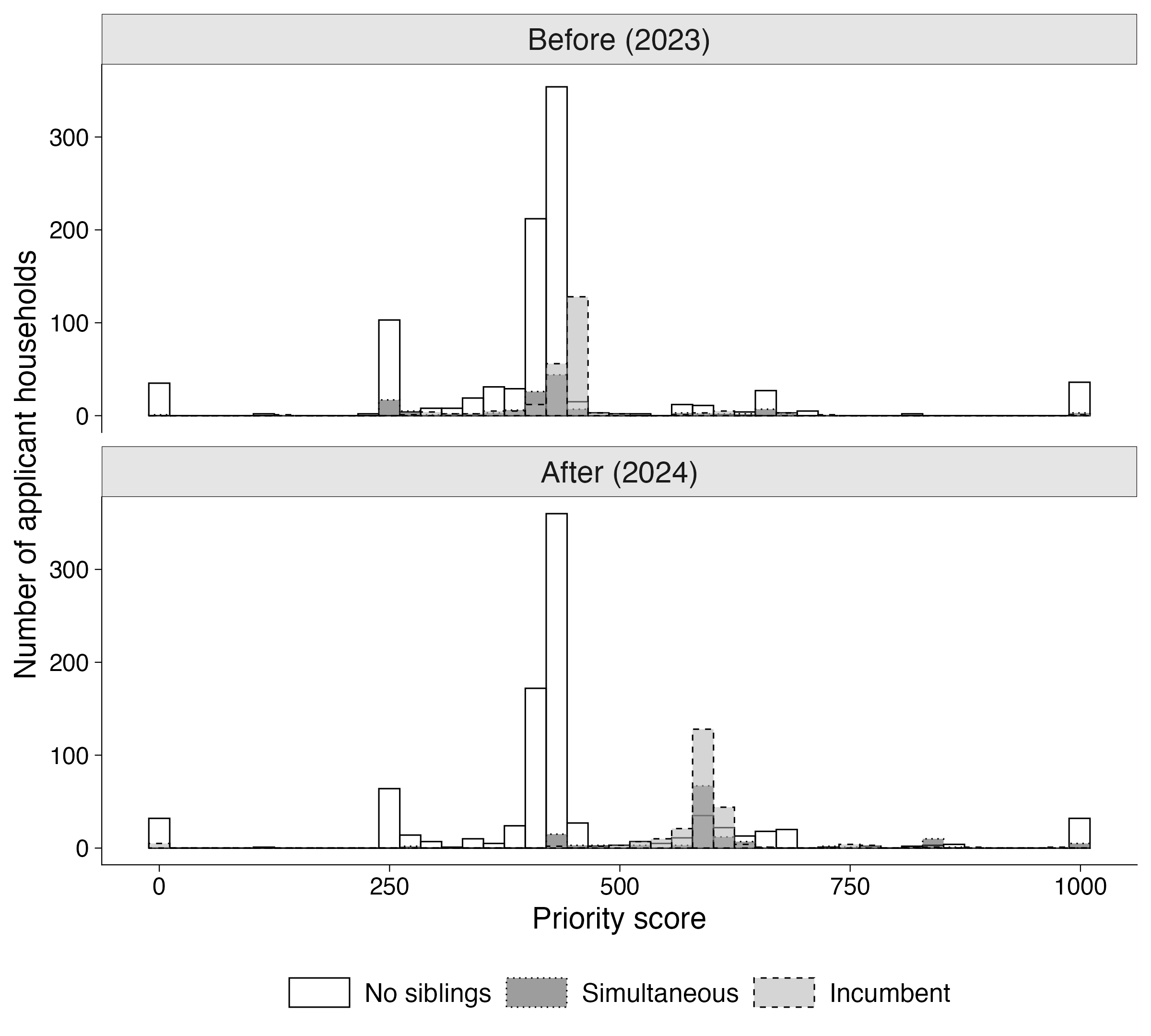}
    \caption{Priority score distributions before and after the 2024 reform}
    \label{fig:score_hist_before_after}
    \begin{minipage}{0.82\linewidth}
        \footnotesize
        \vspace{0.5em}
        \textit{Notes:} The figure plots household-level priority score distributions in the pre-reform year (2023) and the post-reform year (2024), separately by sibling status. Each applicant household contributes one observation. The histograms use 45 bins on a common priority-score scale, and the vertical axis reports the number of applicant households.
    \end{minipage}
\end{figure}

Figure \ref{fig:score_hist_before_after} shows how the distribution of priority scores changed before and after the reform. Before the reform, the three groups were concentrated around a common baseline generated by the scoring formula. A household in which both parents worked at least 140 hours per month received 200 points per parent, while care by a parent on childcare leave added 30 points, producing the prominent mass at 430. Simultaneous applicants received no additional sibling priority and therefore shared this mass with no-sibling applicants. Incumbent applicants were concentrated slightly higher, at 455, because they received an additional 25 points under the pre-reform rule. After the reform, no-sibling applicants remained concentrated at 430, whereas both sibling distributions shifted toward 600. For a representative sibling household, this score combines the 400 employment points, 30 parental-leave points, the new 160-point sibling bonus, and the 10-point multiple-child adjustment. Consequently, a score of 430 fell from the 57th percentile of the aggregate household score distribution in 2023 to the 39th percentile in 2024, illustrating the relative loss of priority faced by no-sibling households.

\subsection{Welfare Evaluation and Priority Design}

To study the welfare impact of changing sibling priority for the fixed 2025 applicant pool and the efficiency--equity tradeoff implied by sibling priority, we conduct a counterfactual welfare analysis using the baseline estimates in Column 1 of Table \ref{tab:estimates}.
Using the 2025 application data, we simulate assignment under alternative policy scenarios and evaluate welfare.\footnote{We focus on the 2025 data because, prior to this year, the municipality relied on a proprietary commercial algorithm provided by a third party, whose assignment outcomes cannot be replicated exactly.}
In these counterfactuals, we hold submitted preferences fixed and vary only the priority structure. Because the assignment mechanism is not guaranteed to be strategy-proof, parental application behavior may itself respond to the presence or absence of sibling priority. As discussed in Section \ref{sec:institution}, households' limited information about the assignment procedure may constrain their ability to adjust reports strategically. A fully rigorous analysis would nevertheless require recomputing reported preferences under the equilibrium that would arise under each policy regime.
For each household, we condition the joint distribution of its random coefficients and tuple-level taste shocks on any ordering that its report implies between the observed 2025 assignment and its simulated pre-reform assignment; Appendix \ref{sec:welfare_conditioning} gives the calculation.
We divide utility by the estimate of $\kappa$ to express welfare in kilometer-equivalent units.

We first compare the observed 2025 assignment with the simulated pre-reform assignment.
Table \ref{tab:welfare-by-sibgroup-conditioned} reports the observed-assignment welfare changes by sibling group under the baseline and individual-choice models, decomposed as in equation \eqref{eq:decomposition_formula}.
Panel A presents the aggregate results from the baseline model. Overall, welfare increases by 0.032 km-equivalent units, from 0.779 under the simulated pre-reform assignment to 0.812 under the observed 2025 assignment. The gain is mainly driven by simultaneous households, whose welfare increases by 1.214 km per household. Households' strong preference for joint assignment already limits the prevalence of split assignment before the reform. Nevertheless, the simulated sibling-priority change further reduces these costs for simultaneous applicants, generating an average welfare gain of 0.039 km through this component. Together with a 1.130 km increase in direct utility and a 0.058 km contribution from preference shocks, partly offset by 0.013 km of additional commuting, this yields their total welfare gain. No-sibling applicants lose 0.089 km. Incumbent applicants lose 0.147 km, mainly because their split-assignment cost and commuting distance rise.

Panel A also shows that changes in the number of children assigned explain only part of the welfare changes. Among simultaneous applicants, the 10 households gaining placements gain 6.737 km on average, while the 117 households with the same number assigned gain 0.772 km. Their benefits therefore reflect not only additional placements but also changes in where children are assigned. Among incumbent applicants, the 7 households gaining placements gain 4.666 km, the 7 losing placements lose 6.002 km, and the remaining 197 households lose 0.110 km on average. Among no-sibling applicants, more households lose placements than gain them (17 versus 8), while the 909 households with the same number assigned lose 0.065 km on average.
Supplemental Appendix Table S5 breaks these effects down by score quartile; simultaneous applicants gain in every quartile, with the largest average gain in the lowest quartile.

Finally, Panel B reports the aggregate results from the individual-choice model. The individual-choice model estimates a slightly larger aggregate gain of 0.034 km but understates the gain for simultaneous applicants by about 27\%, estimating 0.883 rather than 1.214 km.
The difference for simultaneous applicants is concentrated in the direct-utility component: the baseline model estimates an increase of 1.130 km, compared with 0.494 km under the individual-choice model. Treating each child independently can attribute reluctance to accept split assignment to a low valuation of daycare rather than to the cost of separating siblings, changing the estimated value of additional placements. The comparison is consistent with this interpretation, although it also reflects differences in estimated distance sensitivity and in the preference shocks.

\begin{table}[p]
\centering
\caption{Welfare effects of the simulated sibling-priority change}
\label{tab:welfare-by-sibgroup-conditioned}
\scriptsize
\begin{threeparttable}
\begin{tabular*}{\textwidth}{@{\extracolsep{\fill}}lrrrrrrrr}
\toprule
\multicolumn{9}{l}{\textit{Panel A: Baseline model}} \\
\addlinespace[0.25em]
Sibling group & $n$ & Before & After & Change & $\Delta U$ & $\Delta\Gamma$ & $\Delta d$ & $\Delta\varepsilon$ \\
\midrule
All applicants & 1,273 & 0.779 & 0.812 & 0.032 & 0.045 & 0.014 & -0.010 & -0.009 \\
\addlinespace[0.25em]
No-sibling & 934 & 0.395 & 0.306 & -0.089 & -0.097 & 0.000 & -0.029 & -0.021 \\
\quad More children assigned & 8 & -0.685 & 2.427 & 3.112 & 3.187 & 0.000 & 0.950 & 0.875 \\
\quad Same number assigned & 909 & 0.360 & 0.295 & -0.065 & -0.054 & 0.000 & -0.001 & -0.012 \\
\quad Fewer children assigned & 17 & 2.768 & -0.121 & -2.889 & -3.922 & 0.000 & -1.979 & -0.946 \\
\addlinespace[0.25em]
Simultaneous & 128 & 1.673 & 2.887 & 1.214 & 1.130 & -0.039 & 0.013 & 0.058 \\
\quad More children assigned & 10 & 0.923 & 7.660 & 6.737 & 7.576 & 0.461 & 0.774 & 0.396 \\
\quad Same number assigned & 117 & 1.754 & 2.526 & 0.772 & 0.640 & -0.041 & -0.052 & 0.040 \\
\quad Fewer children assigned & 1 & -0.299 & -2.635 & -2.336 & -6.021 & -4.810 & -0.085 & -1.209 \\
\addlinespace[0.25em]
Incumbent & 211 & 1.940 & 1.792 & -0.147 & 0.017 & 0.110 & 0.056 & 0.002 \\
\quad More children assigned & 7 & 3.342 & 8.009 & 4.666 & 4.237 & 0.000 & 0.000 & 0.430 \\
\quad Same number assigned & 197 & 1.669 & 1.558 & -0.110 & 0.012 & 0.077 & 0.043 & -0.002 \\
\quad Fewer children assigned & 7 & 8.163 & 2.161 & -6.002 & -4.065 & 1.144 & 0.479 & -0.314 \\
\bottomrule
\end{tabular*}
\vspace{0.8em}
\begin{tabular*}{\textwidth}{@{\extracolsep{\fill}}lrrrrrrrr}
\toprule
\multicolumn{9}{l}{\textit{Panel B: Individual-choice model}} \\
\addlinespace[0.25em]
Sibling group & $n$ & Before & After & Change & $\Delta U$ & $\Delta\Gamma$ & $\Delta d$ & $\Delta\varepsilon$ \\
\midrule
All applicants & 1,273 & 1.422 & 1.456 & 0.034 & 0.015 & 0.000 & -0.012 & 0.007 \\
No-sibling & 934 & 1.153 & 1.096 & -0.056 & -0.041 & 0.000 & -0.029 & -0.044 \\
Simultaneous & 128 & 3.090 & 3.974 & 0.883 & 0.494 & 0.000 & 0.044 & 0.434 \\
Incumbent & 211 & 1.605 & 1.523 & -0.082 & -0.030 & 0.000 & 0.028 & -0.025 \\
\bottomrule
\end{tabular*}
\begin{tablenotes}[flushleft]
\scriptsize
\item Notes: Panel A reports results from the baseline model by sibling group and decomposes them by changes in licensed placements. Panel B reports aggregate results from the individual-choice model. Before is the assignment simulated under the pre-reform priority rule and After is the observed 2025 assignment. All entries except $n$ are in kilometer-equivalent units. Change equals $\Delta U-\Delta\Gamma-\Delta d+\Delta\varepsilon$. In both models, $\Delta U$ includes the conditional mean contribution of the random coefficients. $\Delta\varepsilon$ is the change in the conditional mean shock: tuple-level in the baseline model and summed over children in the individual-choice model. In Panel A, the same-number category includes unchanged assignments and changes in facilities or in which child is assigned; within each sibling group, weighting the three assignment-change rows by their household shares recovers the sibling-group row. The baseline model conditions on the reported tuple comparison; the individual-choice model conditions on each child's reported comparison. Unlike the policy-grid figures, After here is the observed assignment, including the higher priority for same-age siblings. Welfare levels are not directly comparable across models; compare changes. Changes are calculated from unrounded values; displayed entries may differ slightly when subtracted or summed.
\end{tablenotes}
\end{threeparttable}
\end{table}

\FloatBarrier

\begin{figure}[p]
    \centering
    \begin{minipage}[t]{0.49\linewidth}
        \centering
        \includegraphics[width=\linewidth]{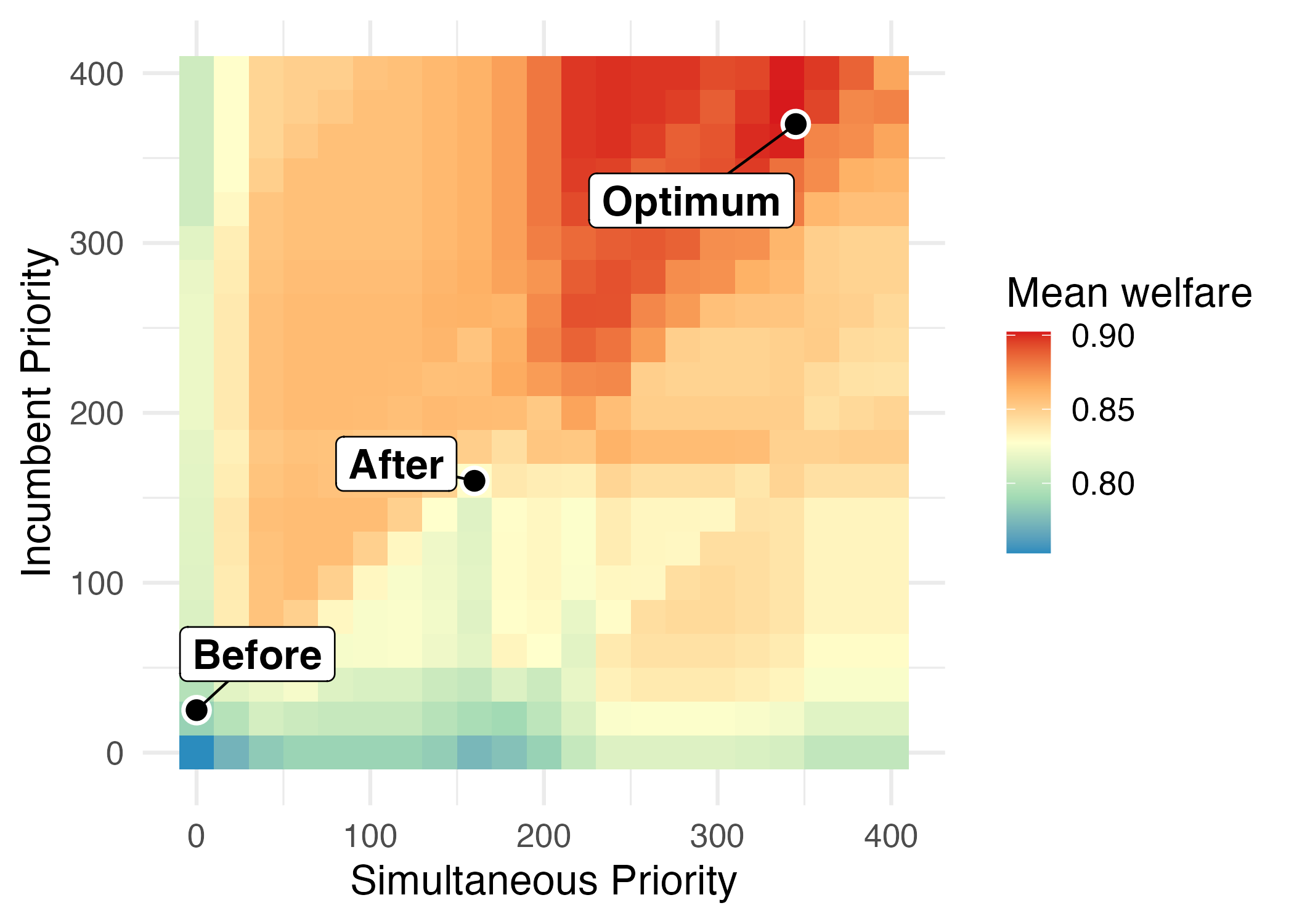}
        \textbf{(a)} Policy grid
    \end{minipage}\hfill
    \begin{minipage}[t]{0.49\linewidth}
        \centering
        \includegraphics[width=\linewidth]{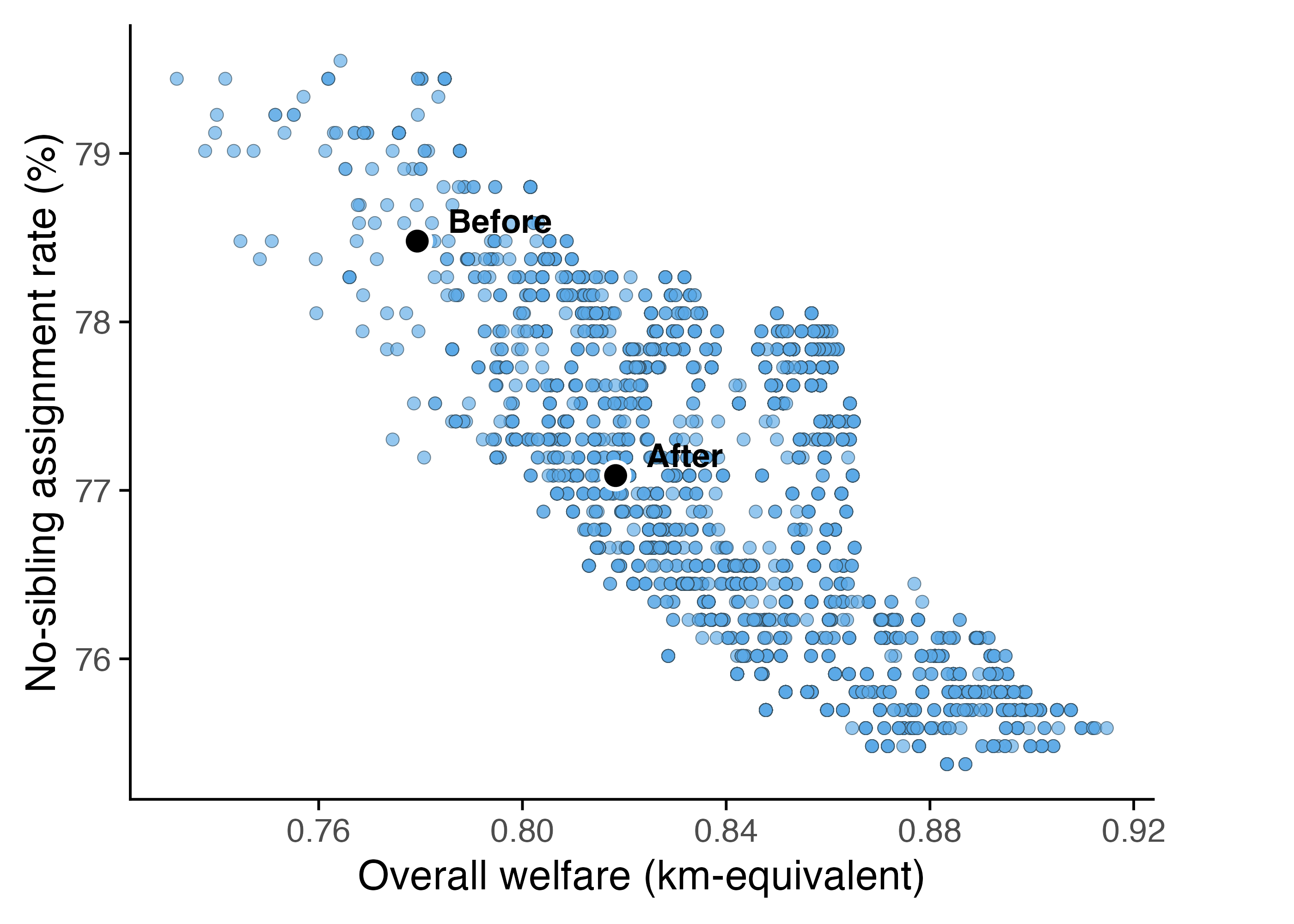}
        \textbf{(b)} Welfare and no-sibling assignment
    \end{minipage}
    \caption{Welfare and sibling-priority policy design}
    \label{fig:policy_grid_and_frontier}
    \begin{minipage}{0.98\linewidth}
        \footnotesize
        \vspace{0.5em}
        \textit{Notes:} Each policy is evaluated using the baseline estimates in Column 1 of Table \ref{tab:estimates}. Panel (a) reports mean welfare over the sibling-priority grid; warmer colors indicate higher welfare. Panel (b) plots mean welfare against the assignment rate among no-sibling applicants. ``Before'' denotes $(0,25)$ and ``After'' denotes $(160,160)$. Welfare is conditional on reported preferences and measured in kilometer-equivalent units.
    \end{minipage}
\end{figure}

To understand the efficiency--equity tradeoff, we next vary the priority points granted to simultaneous applicants, $t_r^{\mathrm{sim}}$, and incumbent applicants, $t_r^{\mathrm{inc}}$, independently on a grid from 0 to 400 in increments of 5. The simulated pre-reform policy (\textit{Before}) corresponds to $(0,25)$, while the simulated post-reform benchmark (\textit{After}) corresponds to $(160,160)$.\footnote{The grid changes only these two sibling-priority bonuses; other components of the 2025 scoring rule are held fixed. The actual reform granted simultaneous applicants 160 points if the siblings were of different ages and 200 points if at least two siblings were of the same age (twins). The grid uses 160 points for both cases; the same-age case is rare in our sample.}
Figure \ref{fig:policy_grid_and_frontier} summarizes the policy-grid results. Panel (a) shows how mean welfare varies across the two priority dimensions. The welfare-maximizing point on the grid is $(t_r^{\mathrm{sim}},t_r^{\mathrm{inc}})=(345,370)$, yielding mean welfare of 0.915, approximately 0.136 kilometer-equivalent units above the pre-reform level. Panel (b) plots mean welfare against the assignment rate for no-sibling applicants, revealing a downward-sloping relationship. The welfare-maximizing policy lowers their assignment rate to 75.6\%, compared with 78.5\% under the pre-reform policy.

\FloatBarrier

\section{Conclusion}
In this paper, we proposed a method to estimate household preferences over joint assignments when siblings apply simultaneously and both placements are determined in the same allocation. A family incurs an additional commuting burden and a fixed cost from split assignment. We translate an extended notion of stability by \citet{sun_stable_2024} into a discrete choice problem, following an approach similar to \citet{fack_beyond_2019}.

We apply our estimation method to the Japanese daycare market, focusing on a municipality in which we assisted in designing a priority reform. For the reference household, we find a non-distance disutility from split assignment equivalent to 4.61 kilometers of commuting, about twice the sample's mean positive commuting distance. Comparing the observed assignment with a counterfactual without the reform that holds reported preferences fixed, we find that welfare gains are concentrated among simultaneous applicants. The individual-choice model understates the gain for this group by about 27\%, while giving a slightly larger aggregate gain.

Our approach is applicable beyond the Japanese daycare context. Existing school choice mechanisms often grant sibling priority, justified by logistical burdens and family considerations, while also raising concerns about equity and diversity. Applying our method allows researchers to quantify the welfare and distributional impacts of such priorities or their absence. This enables a more systematic and evidence-based approach to priority design, which has often been determined through political processes rather than empirical analysis and has largely been treated as given in the market design literature.

Our paper has several limitations. First, our fixed-report counterfactuals hold submitted preferences at their observed values and therefore need not represent an equilibrium of the reporting game under alternative priority policies. Because the mechanism used by the municipality is not guaranteed to be strategy-proof, applicants may strategically adjust their reports in response to changes in the priority system. Second, we do not identify a globally optimal priority structure and instead focus on a two-dimensional sibling-priority design motivated by the relevant policy discussion. Exploring optimal priority design more generally is challenging due to the high-dimensional nature of the problem and remains an important direction for future research.

\section*{Data availability}
The administrative data used in this study were provided by Koriyama City under confidentiality restrictions and are not publicly available. The data contain sensitive municipal administrative information and cannot be shared.

\FloatBarrier
\clearpage
\appendix

\section{Completion of Sibling Preference Reports}\label{sec:preference_completion}

This appendix describes how the assignment system converts the structured application submitted by a sibling household into a complete ordering over assignment tuples. Let $C(f)$ denote the children included in household $f$'s application, let $L_c$ denote child $c$'s ROL, and let $\varnothing$ denote no new placement. For each child, $\varnothing$ means that the child retains their outside option $\overline D(c)$, which need not be parental care; $0$ denotes parental care exclusively. An assignment tuple is $\delta=(\delta(c))_{c\in C(f)}$, where $\delta(c)$ is either a facility on $L_c$ or $\varnothing$.

The household chooses one of four processing rules. The full/partial dimension determines whether partial placement is acceptable, while the joint/split dimension determines whether placement at different facilities is acceptable:

\begin{table}[htbp]
\centering
\small
\begin{tabular}{p{0.25\textwidth}p{0.30\textwidth}p{0.35\textwidth}}
\toprule
Reported rule & Placement requirement & Facility requirement \\
\midrule
Joint/full & All applying children are placed, or none is placed. & All placed children attend the same facility. \\
Joint/partial & A subset of applying children may be placed. & All placed children attend the same facility. \\
Split/full & All applying children are placed, or none is placed. & Different facilities are permitted. \\
Split/partial & A subset of applying children may be placed. & Different facilities are permitted. \\
\bottomrule
\end{tabular}
\caption{Sibling-application processing rules}
\label{tab:sibling_processing_rules}
\end{table}

The completion proceeds in three steps. First, the system enumerates the Cartesian product of the child-level ROLs, adding non-placement as an option for each new applicant. Second, it removes the all-unassigned tuple from the submitted list and removes tuples that violate the selected processing rule. The all-unassigned outcome remains the outside option separating acceptable from unacceptable tuples. Third, it ranks the acceptable tuples lexicographically.

For an acceptable tuple $\delta$, define
\[
W_f(\delta)=\sum_{c\in C(f)}\mathbf{1}\{\delta(c)\neq \varnothing\},
\qquad
S_f(\delta)=\max_{d\in\mathcal D^I}\sum_{c\in C(f)}\mathbf{1}\{\delta(c)=d\},
\]
where $W_f(\delta)$ is the number of children placed and $S_f(\delta)$ is the largest number placed together at one facility. Let $r_c(d)$ be the rank of facility $d$ on child $c$'s ROL, indexing children from oldest to youngest. The implementation sets $r_c(d)=99$ when $d=\varnothing$ or $d$ is not listed, placing it below every facility on an ROL, whose maximum length is ten. The system sorts tuples in descending lexicographic order using
\begin{equation}\label{eq:reported_bundle_completion}
\mathbf{s}_f(\delta)
=
\left(
W_f(\delta),
S_f(\delta),
-\prod_{c\in C(f)}r_c(\delta(c)),
-r_1(\delta(1)),
-r_2(\delta(2)),
\ldots
\right).
\end{equation}
Thus, the completion first favors placing more children, then favors co-location, then minimizes the product of the children's ROL ranks, and finally uses the child-specific ranks from older to younger to break any remaining ties. For a child applying to transfer from a current facility, the status-quo tuple is appended as the final fallback.

For illustration, suppose two new-applicant siblings submit identical lists $L_1=L_2=(A,B)$. The induced ordering is
\begin{align*}
\text{joint/full:}\quad
&(A,A)\succ(B,B)\succ(\varnothing,\varnothing),
\\
\text{joint/partial:}\quad
&(A,A)\succ(B,B)\succ(A,\varnothing)\succ(\varnothing,A)\succ(B,\varnothing)\succ(\varnothing,B)\succ(\varnothing,\varnothing),
\\
\text{split/full:}\quad
&(A,A)\succ(B,B)\succ(A,B)\succ(B,A)\succ(\varnothing,\varnothing),
\\
\text{split/partial:}\quad
&(A,A)\succ(B,B)\succ(A,B)\succ(B,A)\\[-2pt]
&\qquad\succ(A,\varnothing)\succ(\varnothing,A)\succ(B,\varnothing)\succ(\varnothing,B)\succ(\varnothing,\varnothing).
\end{align*}
Tuples omitted from a line violate that processing rule and are treated as unacceptable. This is an administrative completion of the submitted message, not an assumption that the resulting ordering necessarily coincides with the household's latent cardinal utility. The estimation assumption concerns stability with respect to latent utility and is stated in Section~\ref{sec:mechanism_stability}.

\FloatBarrier
\clearpage

\section{Summary Statistics for Daycare Facilities}\label{sec:daycare_sumstats}

Table \ref{tab:daycare_summary} reports summary statistics for licensed childcare facilities in the municipality. Total vacant seats consistently exceed the number of applicants reported in Table \ref{tab:demographics_balance}. In addition to these licensed facilities, the municipality has 32 kindergartens and 31 nonlicensed daycare facilities as of 2025.

\begin{table}[H]
\centering
\linespread{1}\selectfont
\anchoredtablecaption{Licensed childcare facilities by year}{tab:daycare_summary}
\begin{threeparttable}
\centering
\fontsize{8}{9}\selectfont
\begin{tabular}[t]{lllll}
\toprule
Variable & 2022 & 2023 & 2024 & 2025\\
\midrule
Licensed Childcare Facilities & 86 & 89 & 89 & 90\\
Facility Type &  &  &  & \\
\hspace{1em}Licensed Nursery School (hoikusho) & 59 & 59 & 59 & 60\\
\hspace{1em}Small-Scale Daycare & 18 & 19 & 19 & 19\\
\hspace{1em}Certified Child Center & 7 & 8 & 8 & 8\\
\addlinespace
\hspace{1em}Employer-Based Daycare & 2 & 3 & 3 & 3\\
Operator Type &  &  &  & \\
\hspace{1em}For-Profit & 27 & 27 & 27 & 27\\
\hspace{1em}Public & 25 & 25 & 25 & 25\\
\hspace{1em}Other & 13 & 14 & 14 & 15\\
\addlinespace
\hspace{1em}Social Welfare Corp. & 13 & 13 & 13 & 13\\
\hspace{1em}School Corp. & 8 & 10 & 10 & 10\\
Vacant Seats (Mean) &  &  &  & \\
\hspace{1em}Age 0 & 7.2 & 6.95 & 6.48 & 6.14\\
\hspace{1em}Age 1 & 6.39 & 5.7 & 5.33 & 5.32\\
\addlinespace
\hspace{1em}Age 2 & 3 & 3.36 & 2.85 & 3.5\\
\hspace{1em}Age 3 & 4.66 & 4.58 & 4.27 & 4.39\\
\hspace{1em}Age 4 & 5.77 & 3.1 & 2.61 & 2.96\\
\hspace{1em}Age 5 & 4.23 & 4.6 & 2.22 & 2.67\\
\hspace{1em}Total Vacant Seats & 1874 & 1756 & 1417 & 1518\\
\addlinespace
Cutoff (Mean) &  &  &  & \\
\hspace{1em}Age 0 & 393.2 & 363.6 & 391.37 & 408.23\\
\hspace{1em}Age 1 & 394 & 399.27 & 463.5 & 445.76\\
\hspace{1em}Age 2 & 342.57 & 442.54 & 580.97 & 470.97\\
\hspace{1em}Age 3 & 445.06 & 515.9 & 611.43 & 591.72\\
\addlinespace
\hspace{1em}Age 4 & 303 & 407 & 605 & 423.67\\
\hspace{1em}Age 5 & 430 & 516.43 & 659.71 & 488\\
\hspace{1em}Share of Positive Cutoffs & 0.182 & 0.507 & 0.545 & 0.527\\
\bottomrule
\end{tabular}
\begin{tablenotes}[flushleft]
\scriptsize
\item Notes: The table summarizes licensed childcare facilities by year. Facility type and operator type rows report counts of distinct facilities. Age-specific mean vacant seats are calculated only among facilities with positive vacancies at that age, so the number of facilities varies by age and year. Total vacant seats is the citywide sum across facility--age cells, not the sum of these conditional means. Mean cutoffs are computed conditional on the cutoff being positive. ``Share of Positive Cutoffs'' is the fraction of facility--age categories for which admission is constrained by a binding cutoff.
\end{tablenotes}
\end{threeparttable}
\end{table}

\FloatBarrier
\clearpage
\section{Estimates of Facility Fixed Effects}\label{sec:alphad_estimates}

\begin{center}
\includegraphics[width=0.68\linewidth]{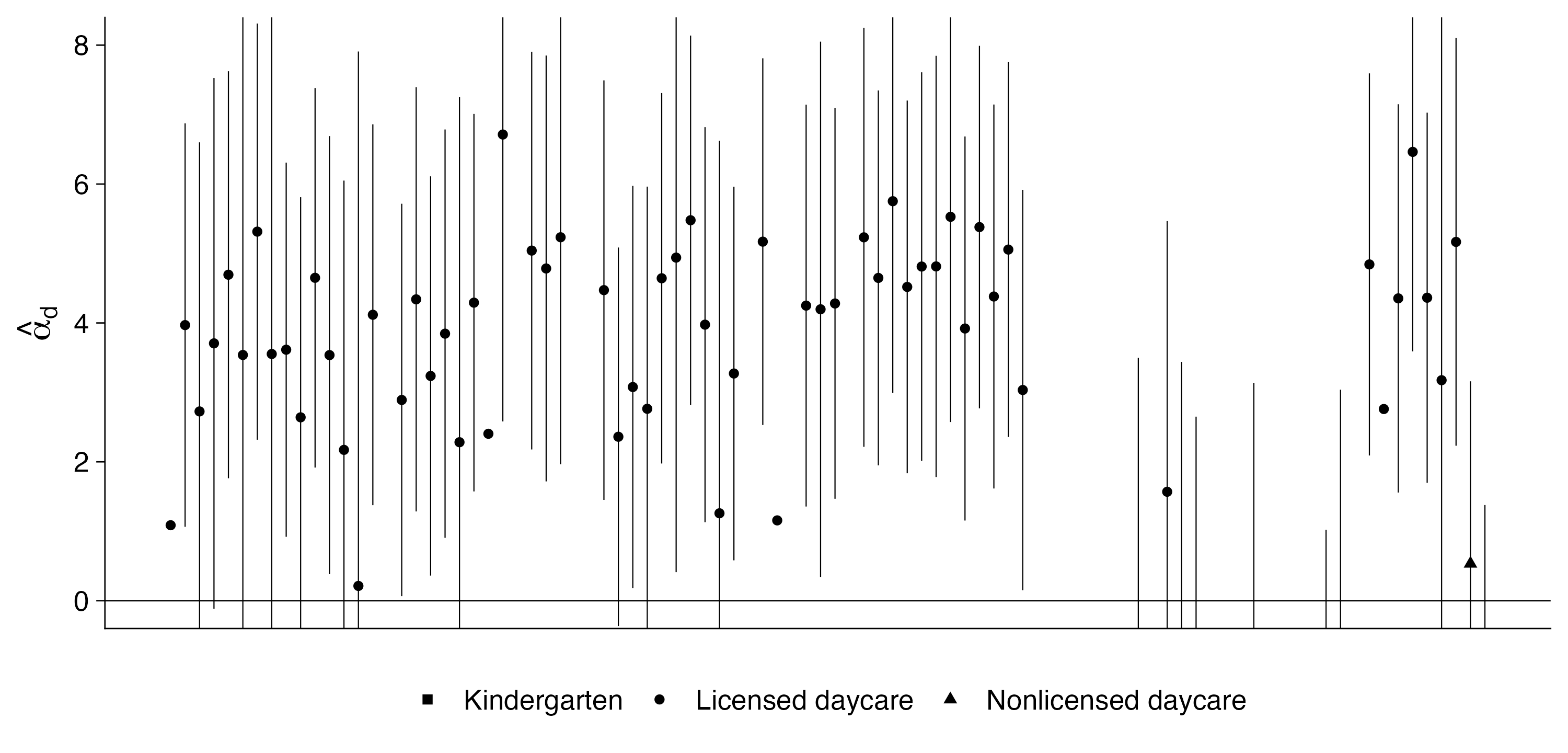}
\anchoredfigurecaption{Facility fixed effects: baseline model}{fig:alphaD-full}
\begin{minipage}{0.68\linewidth}
\scriptsize
\emph{Notes:} The figure plots estimated facility fixed effects $\hat{\alpha}_d$ from the baseline model. Points correspond to point estimates, and vertical bars indicate 95\% confidence intervals. Confidence intervals are omitted for 26 fixed effects with standard errors greater than 4; their point estimates remain plotted. Licensed childcare facilities are shown individually, while nonlicensed daycare and kindergarten are each represented by a single fixed effect.
\end{minipage}
\end{center}

\begin{center}
\includegraphics[width=0.68\linewidth]{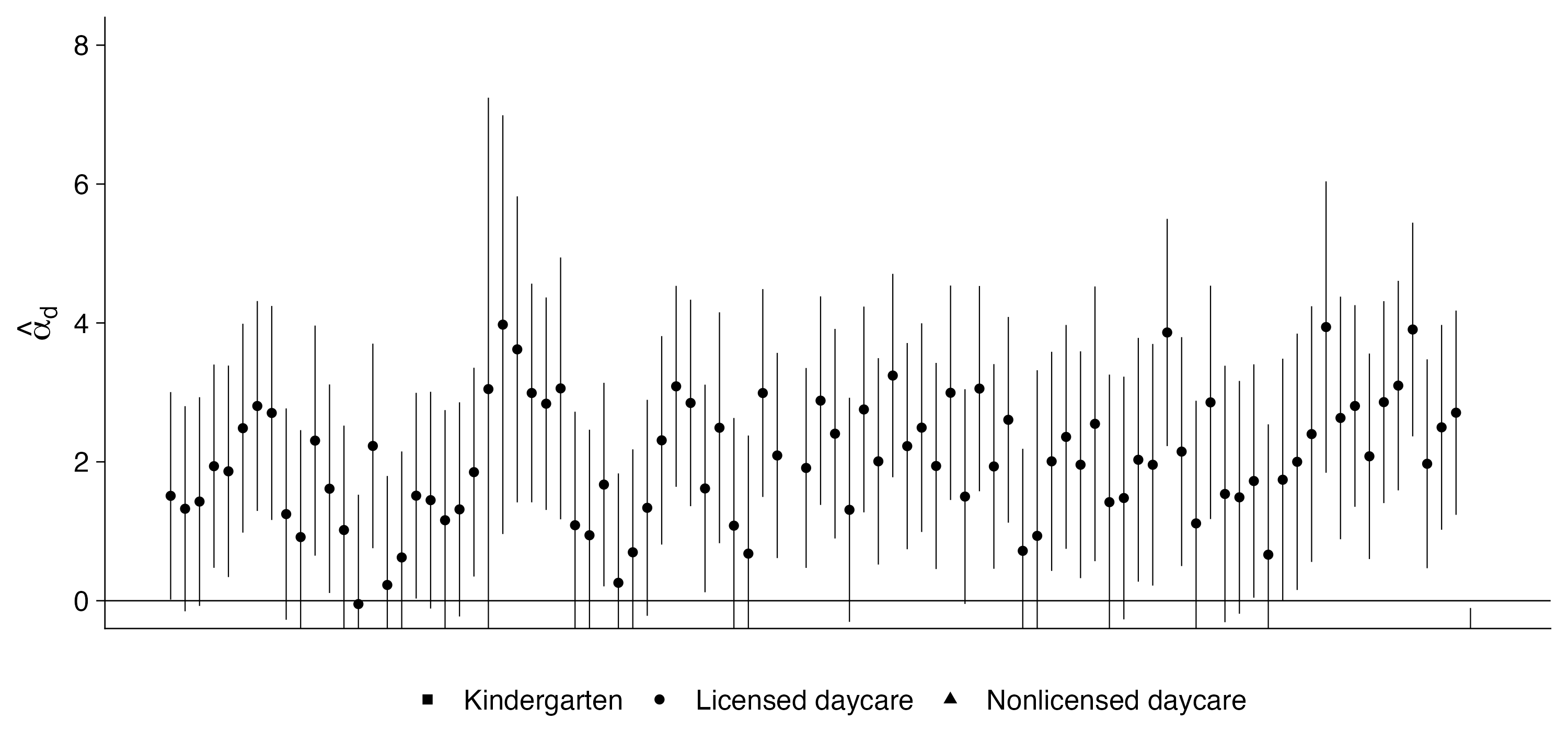}
\anchoredfigurecaption{Facility fixed effects: individual-choice model}{fig:alphaD-ind}
\begin{minipage}{0.68\linewidth}
\scriptsize
\emph{Notes:} The figure plots estimated facility fixed effects $\hat{\alpha}_d$ from the individual-choice model without sibling complementarities. Points correspond to point estimates, and vertical bars indicate 95\% confidence intervals. All plotted fixed effects have standard errors at or below 4, so no confidence intervals are omitted. Licensed childcare facilities are shown individually, while nonlicensed daycare and kindergarten are each represented by a single fixed effect.
\end{minipage}
\end{center}

\FloatBarrier
\clearpage
\FloatBarrier
\clearpage
\section{Welfare Conditioning on Reported Preferences}\label{sec:welfare_conditioning}

Suppose family $f$ receives assignment $\delta^r$ under policy scenario $r$. Let $\delta^{Before}$ be its assignment simulated under the pre-reform rule, and let $\delta^{After}$ be its observed 2025 assignment. We condition on the ordering, if any, that the household's report implies between these two assignments. For this welfare calculation, we treat that comparison as a restriction on latent utility; we do not impose the report's other comparisons. The same ordering is used to evaluate every policy scenario; we do not condition on the fact that $\delta^{After}$ was assigned. For the derivation, suppose the report implies $\delta^{After}\succ_f\delta^{Before}$. The reverse case exchanges the two assignments.

We evaluate family $f$'s welfare under policy $r$ as $\E[u_{f,\delta^r}\mid\delta^{After}\succ_f\delta^{Before}]$. From the utility specification in equation \eqref{eq:decomposition_formula},
\[
u_{f,\delta^r}=V_{f,\delta^r}(\nu_f;\hat\vartheta)+\varepsilon_{f,\delta^r},
\]
where $\nu_f$ has a standard bivariate normal distribution and the tuple shocks independently follow a unit-scale Type I extreme-value distribution. We draw $\nu_f^{(m)}$ from this normal distribution for $m=1,\dotsc,M$. Conditional on $\nu_f$, the probability of the reported ordering is
\[
\Pr(\delta^{After}\succ_f\delta^{Before}\mid\nu_f)
=
\frac{\exp\{V_{f,\delta^{After}}(\nu_f;\hat\vartheta)\}}
{\exp\{V_{f,\delta^{After}}(\nu_f;\hat\vartheta)\}
+\exp\{V_{f,\delta^{Before}}(\nu_f;\hat\vartheta)\}}.
\]
Bayes' rule gives the conditional density of the random coefficients:
\[
p(\nu_f\mid\delta^{After}\succ_f\delta^{Before})
=
\frac{\Pr(\delta^{After}\succ_f\delta^{Before}\mid\nu_f)\,\phi(\nu_f)}
{\int\Pr(\delta^{After}\succ_f\delta^{Before}\mid\nu')\,\phi(\nu')\,d\nu'}.
\]
Since the draws already come from the normal prior $\phi$, we implement this update by assigning draw $m$ the weight
\[
w_{fm}
=
\frac{\Pr(\delta^{After}\succ_f\delta^{Before}\mid\nu_f^{(m)})}
{\sum_{\ell=1}^{M}\Pr(\delta^{After}\succ_f\delta^{Before}\mid\nu_f^{(\ell)})}.
\]

The same reported ordering also informs the tuple shocks. Write $p_{fm}=\Pr(\delta^{After}\succ_f\delta^{Before}\mid\nu_f^{(m)})$. Conditional on draw $m$, the shock entering welfare under policy $r$ has mean
\[
\E[\varepsilon_{f,\delta^r}\mid\delta^{After}\succ_f\delta^{Before},\nu_f^{(m)}]
=
\begin{cases}
\gamma_{\mathrm E}-\log p_{fm},
& \delta^r=\delta^{After},\\[3pt]
\gamma_{\mathrm E}+\dfrac{1-p_{fm}}{p_{fm}}\log(1-p_{fm}),
& \delta^r=\delta^{Before},\\[6pt]
\gamma_{\mathrm E},
& \delta^r\notin\{\delta^{After},\delta^{Before}\},
\end{cases}
\]
where $\gamma_{\mathrm E}$ is the mean of a unit-scale Type I extreme-value shock.\footnote{Fix $\nu_f^{(m)}$ and suppress the argument $(\nu_f^{(m)};\hat\vartheta)$ of $V_{f,\delta}$ in this footnote. The event $\delta^{After}\succ_f\delta^{Before}$ requires $\varepsilon_{f,\delta^{After}}-\varepsilon_{f,\delta^{Before}}>V_{f,\delta^{Before}}-V_{f,\delta^{After}}$. Integrating the EV1 density of $\varepsilon_{f,\delta^{After}}$ against the probability that the other shock satisfies this inequality gives
\begin{align*}
&\E[\varepsilon_{f,\delta^{After}}\mid\delta^{After}\succ_f\delta^{Before},\nu_f^{(m)}]\\
&\quad=\frac{1}{p_{fm}}\int_{-\infty}^{\infty}
z e^{-z}\exp\!\left[-\left(1+e^{V_{f,\delta^{Before}}-V_{f,\delta^{After}}}\right)e^{-z}\right]dz\\
&\quad=\gamma_{\mathrm E}-\log p_{fm},
\end{align*}
using $p_{fm}=\bigl(1+e^{V_{f,\delta^{Before}}-V_{f,\delta^{After}}}\bigr)^{-1}$. For the other shock, partition its unconditional mean according to which assignment is preferred:
\begin{align*}
\gamma_{\mathrm E}
&=p_{fm}\E[\varepsilon_{f,\delta^{Before}}\mid\delta^{After}\succ_f\delta^{Before},\nu_f^{(m)}]\\
&\quad+(1-p_{fm})\E[\varepsilon_{f,\delta^{Before}}\mid\delta^{Before}\succ_f\delta^{After},\nu_f^{(m)}].
\end{align*}
The final expectation equals $\gamma_{\mathrm E}-\log(1-p_{fm})$ by the same calculation with the assignments exchanged. Solving gives the second case.} Thus, the reported comparison updates the shocks to the two compared assignments, while a different tuple's shock retains its unconditional mean.

Finally, family $f$'s expected utility under policy $r$ is calculated as
\[
\E[u_{f,\delta^r}\mid\delta^{After}\succ_f\delta^{Before}]
\approx
\sum_{m=1}^{M}w_{fm}
\left[
V_{f,\delta^r}(\nu_f^{(m)};\hat\vartheta)
+\E[\varepsilon_{f,\delta^r}\mid\delta^{After}\succ_f\delta^{Before},\nu_f^{(m)}]
\right].
\]
The weights are formed once per household and applied to the entire utility at each draw. If the report implies $\delta^{Before}\succ_f\delta^{After}$, we use that ordering in the same calculation. If the assignments coincide or the report reveals no ordering between them, there is no update: $w_{fm}=1/M$ and the tuple shock has mean $\gamma_{\mathrm E}$. We divide expected utility by the estimated distance coefficient $\hat\kappa$ to express welfare in kilometer-equivalent units, then average across households.

For the individual-choice model, we apply the same calculation separately to each child, conditioning that child's independent random coefficients and facility shocks on the reported ordering between their Before and After facilities. A listed facility is preferred to the child's outside option. If the facilities coincide or the report does not order them, that child's distribution is not updated. Each child's posterior is held fixed across policy scenarios, and the resulting expected utilities are summed within the original household before averaging across households. The sibling-processing rule does not enter these child-level comparisons.

\clearpage

\FloatBarrier
\clearpage
\bibliographystyle{plainnat}
\bibliography{references}

\end{document}